\documentclass[sigconf]{acmart}
\usepackage{subcaption}
\usepackage{soul} 
\sethlcolor{yellow}
\usepackage[table,xcdraw]{xcolor}
\usepackage{minted}

\AtBeginDocument{%
  }

\setcopyright{acmlicensed}
\copyrightyear{2024}
\acmYear{2024}

\begin{document}

\title{DepthScape: Authoring 2.5D Designs via Depth Estimation, Semantic Understanding, and Geometry Extraction}

\author{Xia Su}
\affiliation{%
  \institution{University of Washington}
  \city{Seattle}
  \state{Washington}
  \country{USA}
}
\email{xiasu@cs.washington.edu}

\author{Cuong Nguyen}
\affiliation{%
  \institution{Adobe Research}
  \city{San Francisco}
  \state{California}
  \country{USA}
}
\email{cunguyen@adobe.com}

\author{Matheus A. Gadelha}
\affiliation{%
  \institution{Adobe Research}
  \city{San Jose}
  \state{California}
  \country{USA}
}
\email{gadelha@adobe.com}

\author{Jon E. Froehlich}
\affiliation{%
  \institution{University of Washington}
  \city{Seattle}
  \state{Washington}
  \country{USA}
}
\email{jonf@cs.washington.edu}

\renewcommand{\shortauthors}{Su et al.}

\begin{abstract}
2.5D effects, such as occlusion and perspective foreshortening, enhance visual dynamics and realism by incorporating 3D depth cues into 2D designs. However, creating such effects remains challenging and labor-intensive due to the complexity of depth perception. We introduce DepthScape, a human-AI collaborative system that facilitates 2.5D effect creation by directly placing design elements into 3D reconstructions. Using monocular depth reconstruction, DepthScape transforms images into 3D reconstructions, where visual contents are placed to automatically achieve realistic occlusion and perspective foreshortening. To further simplify 3D placement through a 2D viewport, DepthScape employs a vision-language model to analyze source images, extracting key visual components as content anchors to enable direct manipulation editing. We evaluate DepthScape among nine participants with varying design skills, confirming the effectiveness of the creation pipeline. We also test on 100 professional stock images to assess robustness, complemented by an expert evaluation that confirms the quality of DepthScape’s results.
\end{abstract}



\begin{teaserfigure}
  \includegraphics[width=\textwidth]{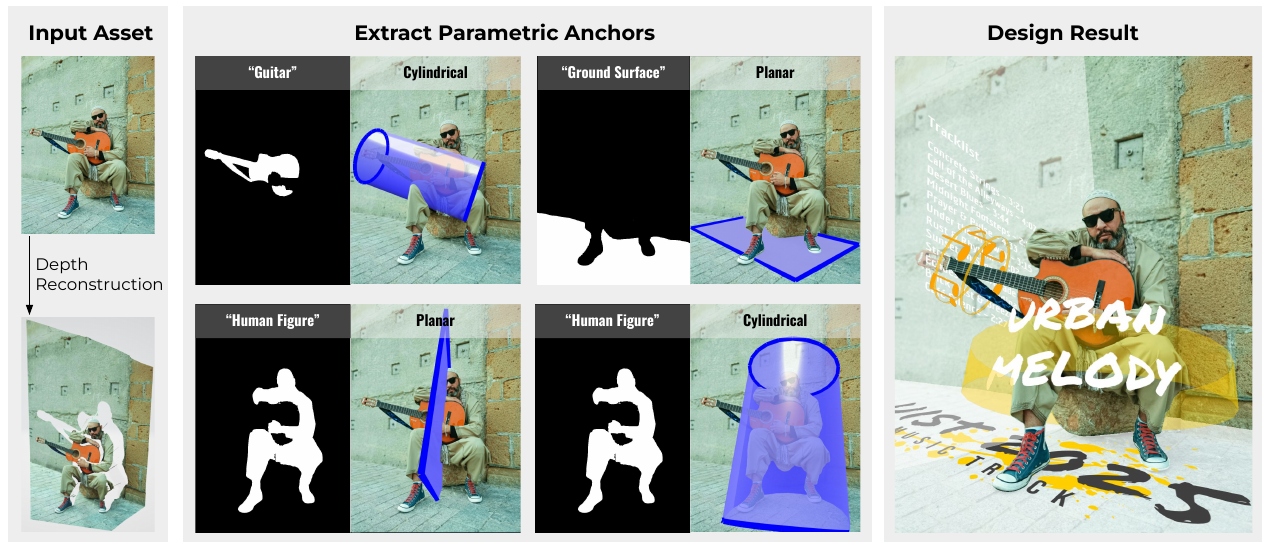}
  \caption{We introduce \textit{DepthScape}, a Human-AI collaborative authoring system for 2.5D visual design. DepthScape takes input image assets and uses 3D reconstruction to estimate its inherent depth information. With AI-assisted design recommendation, users can quickly  layout design elements in the implicit 3D space. The output is a visual design with realistic occlusion effects following depth cues in the input image.}
  \Description{From left to right shows the pipeline of DepthScape. Top left shows an input asset image, which is a musician figure holding a guitar, while sitting at a wall corner. Below it is its 3D reconstruction, showing the depth information of this image. To the right are four extracted parametric anchors, showing a blue cylinder surface around the guitar, plane on the ground under the figure's feet, plane in the middle of the figure, and a cylinder around the figure. Right shows the design results, which is an album cover with a surrounding ring of text saying "urban melody", while text on the ground says "UIST 2025 Music Track". A ring of music notes surrounds the top of the guitar, and a half-transparent white surface containing text of the tracklist is placed in the background. }
  \label{fig:teaser}
\end{teaserfigure}

\maketitle


\section{Introduction}
\begin{figure*}
    \centering
    \includegraphics[width=1\linewidth]{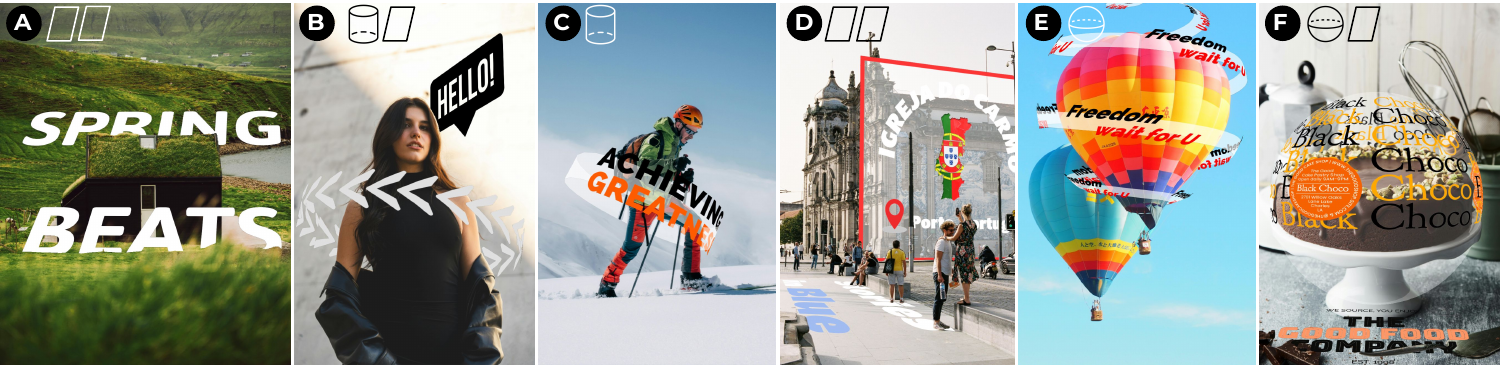}
    \caption{Design result gallery of DepthScape. Top row icons show the formation of designs, with one or multiple Planar, Cylindrical, and Spherical parametric anchors.}
    \label{fig:gallery}
    \Description{A list of design results. From left to right: a grassland view with a little house in it, with text showing "SPRING BEATS"; A woman standing in front of a cement wall, with a gray arrow circle surrounding her and a text plane ""Hello" on her right side; A hiking man walking with great effort on a snowy ground, with a text ring surrounding him saying "ACHIEVING GREATNESS"; An European street scene with a church in scene and some tourists on the foreground. A plane with the church name is placed on the church wall, and a plane with "Stories in Blue" is placed in parallel with the ground; A hot air balloon scene, with the front red balloon surrounded by a spiral text saying "Freedom wait for U"; A table with a cake on it. The cake is wrapped with a sphere saying "Black Chocolate" in a repetition. There's also a text saying "THE GOOD FOOD COMPANY" on the tale surface. }
\end{figure*}

Humans gain 3D perception from 2D images through depth cues like occlusion, perspective, and shading \cite{murray1994some,brenner2018depth,howard2002depth}. Leveraging this visual ability, creators have long applied depth-enhancing techniques such as layering and projecting to achieve visual realism. Borrowing from graphics research, we call this type of visual design ``\textit{2.5D design}'', whose perceptual power has long been recognized in Gestalt principles \cite{pylyshyn1999vision}, visual perception research \cite{man1982computational,wu2017marrnet}, and design practice. However, creating 2.5D visual designs can be challenging.

Consider Jennifer, a 2D visual designer trying to make a design poster more engaging. Jennifer aims to add 3D effects like a decoration wrap around the model's body, and a 3D text box near the model's face (\autoref{fig:gallery}BC). Simulating 3D effects using pure 2D layer-based design tools like Photoshop or Illustrator is tedious and error-prone. Even minor adjustments, such as rotation or repositioning, demand repeated refinements and strong spatial intuition, making iteration slow and frustrating. Jennifer might consider using a 3D tool like Blender. She could segment the main object ~\cite{ravi2024sam2segmentimages}, set up a 3D scene, and block out simple 3D geometry to get accurate depth cues and vanishing points. However, 3D tools have steep learning curves and introduce additional challenges like ensuring the 3D geometries align well with the existing photograph. These tools can be overwhelming for 2D designers who might not be willing to invest time in mastering 3D workflow. In summary, both workflows shift attention from creative exploration to technical problem-solving, rendering the process tedious, error-prone, and resistant to rapid experimentation.

\begin{figure}
    \centering
    \includegraphics[width=0.6\linewidth]{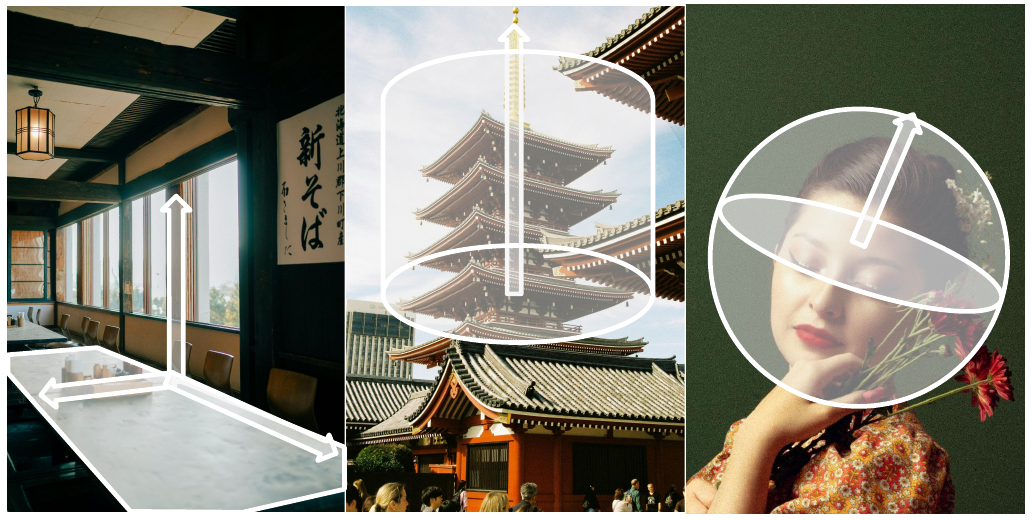}
    \caption{Three types of parametric anchoring supported by DepthScape. From left to right: Planar, Cylindrical, Spherical}
    \label{fig:types}
    \Description{From left to right: a Japanese restaurant scene, with a rectangle table in the middle of the scene. There is a half-transparent white surface on top of the table, with some arrows showing the XYZ direction of this plane; A Japanese temple scene with a tower in the center of the photo. A white half-transparent cylinder surrounds the tower, with an upward arrow indicating the central axis; A woman holding some flowers. A half-transparent white sphere surrounds her head.}
\end{figure}

A more streamlined approach is to allow 2D designers to simply drop assets into an image and watch them instantly adapt to the right 3D perspective and occlusion of the photograph. We envision a human–AI collaborative approach in which designers remain on the 2D canvas while an AI agent handles complex 3D tasks in the background such as reconstruction, projection, and rendering. The AI agent automatically analyzes the RGB content and depth maps of an image and anchors perspective-correct 3D primitives like planes, cylinders, or spheres on objects in the image space. These ``\textit{parametric anchors}'' enable 2D assets to be placed with correct perspective and occlusion. \autoref{fig:types} show examples of parametric anchors. Users can then drop design assets on these anchors to quickly get 2.5D effects, with the option to refine the result through direct manipulation and detailed parameter configuration. This interaction resembles an offline form of augmented reality scene creation~\cite{leiva2020pronto}, allowing designers to author 2.5D effects without mastering traditional 3D modeling.

To realize this vision, we developed \textit{DepthScape}, a web-based tool for simplified 2.5D design. Our goal is to enable both professional and amateur 2D designers to create sophisticated 2.5D effects without adopting a complex 3D workflow. A key contribution of DepthScape is the use of a vision–language model (VLM), GPT-4o~\cite{hurst2024gpt}, to orchestrate an RGB–and–depth processing pipeline that automatically generates parametric anchors. We leverage two core capabilities of the VLM—visual reasoning and code generation—to analyze the input image and synthesize custom geometry extraction programs (see \autoref{fig:extraction}A). For example, for the model image in \autoref{fig:gallery}B, DepthScape may produce two programs: one that fits a 3D cylinder to support a wrapping effect around the body, and another that extracts a 3D surface oriented along the model’s gaze direction. Each program (\autoref{fig:extraction}A) comprises a sequence of connected function calls that perform tasks such as masking, point-cloud cleaning, primitive fitting, and human pose or face analysis. We implement these functions as a reusable library and provide in-context examples to guide the VLM during code generation.

A key benefit of this \textit{visual programming} approach~\cite{gupta2023visual} is scalability and adaptability: the system can automatically generate diverse types of parametric anchors based on image content. These anchors facilitate 2.5D design by (i) offering recommendations that help novice designers explore what is possible, and (ii) enabling experienced designers to rapidly try alternatives. \autoref{fig:gallery} showcases the range of effects created by users with our system.


The design of DepthScape was informed by a human-centered, iterative design process. We initially started with a proof-of-concept prototype to test the feasibility of our core idea: rendering 2.5D visual designs by placing visual elements into reconstructed depth spaces. We conducted a user study with nine participants to evaluate usability, explore potential usage scenarios, and gather feedback on interaction methods. Participants replicated example 2.5D designs and created open-ended designs, confirming the feasibility of our approach while also uncovering desired user interactions and system capabilities. These insights directly support the usability of the system for both professional and amateur users, and informed the development of a second prototype, which improved perspective rendering, added direct manipulation widgets, and enhanced AI assistance.

To evaluate the final DepthScape prototype, we performed a technical performance evaluation with 100 professional stock photos of diverse types and content. By feeding these images into DepthScape and logging the outputs, we evaluate the efficiency, robustness, and output diversity. 
To further evaluate the quality of outputs, we also conducted an expert evaluation session with three professional designers, who positively rated the quality of suggested anchors and design outputs. To further demonstrate DepthScape's potential and flexibility, we additionally explore five application scenarios including 2.5D video creation, real-world scene modification, and also storyboarding for hand-sketched scenes.



Our contribution is threefold. First, we propose a novel authoring pipeline for 2.5D visual designs by placing visual elements into monocular depth reconstructions of input images. Second, we employ a VLM and visual programs to automatically extract parametric anchors from input images based on image semantics, which further enables direct-manipulative editing. Third, we demonstrate the flexibility, robustness, and potential of our approach via our pilot user studies, technical performance evaluations with expert review, and five application examples. 



\section{Related Work}
We situate our contributions in literature on 2.5D design, depth-aware design tools, and semantic geometry extraction.

\subsection{2.5D design}
Humans gain 3D perception not only from binocular vision but also monocular cues to depth, such as linear perspective, interposition, Gestalt principles, and shadows \cite{murray1994some,brenner2018depth,howard2002depth}. Since such depth perception comes from how people mentally process visual information, some researchers also call them \textit{psychological} depth cues \cite{mehrabi2013making}.
Techniques based on this instinctive human ability are widely applied in graphics research and engineering. From the 1970s, developers have been using methods like scaling sprites \cite{giantbomb_sprite_scaling,giantbomb_fonz,giantbomb_pole_position,giantbomb_turbo} and parallax scrolling \cite{giantbomb_parallax_scrolling} to create pseudo-3D effects in arcade games. Although hard to identify the exact origin, the term \textit{2.5D} gradually emerged from the animation and gaming community back to 1970s and 1980s to refer to the pseudo-3D effects seen in the 2D visuals.

Similar techniques are also widely applied in visual arts from paintings --- artists like Rembrandt manifest full working knowledge of all monocular cues to depth in their pieces back to 17th century \cite{howard2002depth}--- to modern designs of posters \cite{iskin2014poster}, illustrations \cite{meggs2025graphic}, websites, \textit{etc}.
The development of modern creation tools like Adobe PhotoShop and Adobe Illustrator further simplifies and supports the creation of visual depth cues thanks to their layering and distorting capability. This further blurs the line between 2D and 3D design spaces. To our knowledge, there isn't a clear definition for these depth-aware designs either as a design genre or a certain technique. In this case, we borrow from the gaming community, graphics research community \cite{wu2015research,6189341,gois2015interactive,narayan2005study,coutinho2016puppeteering,fukusato2021view}, and recent creativity support \cite{zhou2024portalink} and VR \cite{10298000} research to address this visual design type as \textit{2.5D design}.

\subsection{Depth-aware Design Tools}
In addition to using design techniques and graphic technologies to achieve 2.5D visual perception, researchers have also increasingly leveraged real estimated depth to assist creative tasks. With recent advancements in computer vision making monocular depth estimation more accurate and efficient, the depth information of a single image can now be easily obtained, whether as a depth map \cite{yang2024depth,yang2024depthV2,Patni2024ECoDepthEC,Zhu2024HABinsHA} or a 3D model \cite{wang2024crm,xu2024grm,xu2024instantmesh,zhao2024flexidreamer,wang2024moge}. These output results can be applied in various creative applications.
For example, \textit{ZoomShop} \cite{liu2022zoomshop} use image depth information to edit image composition, enlarge distant objects, and adjust the relative size and positions of objects.
Similarly, \textit{VideoDoodle} \cite{10.1145/3592413} uses depth information of video scenes to blend hand-drawn animations into real-world video scenes.
Lu \textit{et al.}\cite{lu2019depth} uses depth maps to vectorize images and better capture key contours in the input image.

In addition to using real or estimated depth information, existing research also enables creators to arrange 2D elements in 3D spaces to better support their creation. The \textit{Mental Canvas} \cite{dorsey2007mental} allows creators to organize 2D architecture sketches into 3D space so that 3D strokes that ensure geometric consistency can be better created; \textit{Stereoboard} \cite{henrikson2016storeoboard} also allows users to arrange 2D storyboard sketches in 3D space so that they can better align with cinematic constructs. \textit{PortalInk} \cite{zhou2024portalink} automatically arranges depth-based layers of 2D drawings into 3D spaces to create parallax effects and export 2.5D visual stories. By organizing 2D contents in 3D spaces, these tools support easy creation of contents that follow 3D depth cues and spatial relations, like architectural structures and 2.5D animations. 

Following this thread of work, we gain insights on how depth information helps creation tasks: depth information provides spatial details that ensure content blending true to reality. This inspires our work, since 2.5D visual designs are also heavily dependent on depth perception and are aimed at visual realism.

\subsection{Semantic Geometry Extraction}
To better understand reconstructed 3D scenes and support creativity, we aim to automatically extract geometry structure from the reconstructed depth scenes. Traditional techniques in geometry fitting like RANSAC \cite{fischler1981random} and tools such as PyRANSAC-3D \cite{mariga_pyRANSAC3D} are commonly used to fit geometric primitives (\textit{e.g.}, planes, spheres, cylinders) to noisy point cloud data. These methods are often paired with supervised segmentation models to extract meaningful structures from real-world scans \cite{kyriakaki2022geometric,8977383,7822913}. As recent advancements in vision-language models (VLMs) \cite{zhang2024vision,radford2021learningtransferablevisualmodels} have enabled AI systems to understand visual content in a zero-shot way, it's now possible to conduct such geometry extraction in zero-shot by using VLM to parse the scenes. For example, using zero-shot text-based segmentation like Grounded-SAM \cite{ren2024grounded}, which enables text-based segmentation in 2D input images, we can automate the selection of certain objects in reconstructed point clouds, which can be subsequently fitted with primitive shapes. Combined with other CV models like MediaPipe \cite{lugaresi2019mediapipe} which detects human skeletons and landmarks, we can further expand this detection capability to match wider design intents.

Inspired by Visual Programming \cite{gupta2023visual}, which demonstrates how large language models (LLMs) can orchestrate modular computer vision tasks by generating visual programs that combine off-the-shelf computer vision modules, we explore a hybrid method to semantically extract geometric primitives from the reconstructed depth spaces. Our pipeline combines VLM-based semantic grounding and segmenting with classical geometric fitting, aiming for a zero-shot, modular approach to identify planar, cylindrical, and spherical structures that can be content anchors that accelerate the creation of 2.5D effects. This enables users to manipulate design elements in a depth-aware space while maintaining intuitive semantic control.




\section{Designing DepthScape}

Building upon the core idea of placing visual contents into reconstructed depth spaces, we conducted a three-stage iterative process to test its feasibility, as well as explore user needs and key interactions. We first built a \textit{proof-of-concept prototype} to test out feasibility. Then, we used our prototype to conduct a \textit{user study} with nine participants with varying design expertise. Finally, based on study findings, we built the \textit{final DepthScape system} with improved depth reconstruction, user interaction, and AI design suggestions.

\subsection{Proof-of-Concept Prototype}
\begin{figure}
    \centering
    \includegraphics[width=\linewidth]{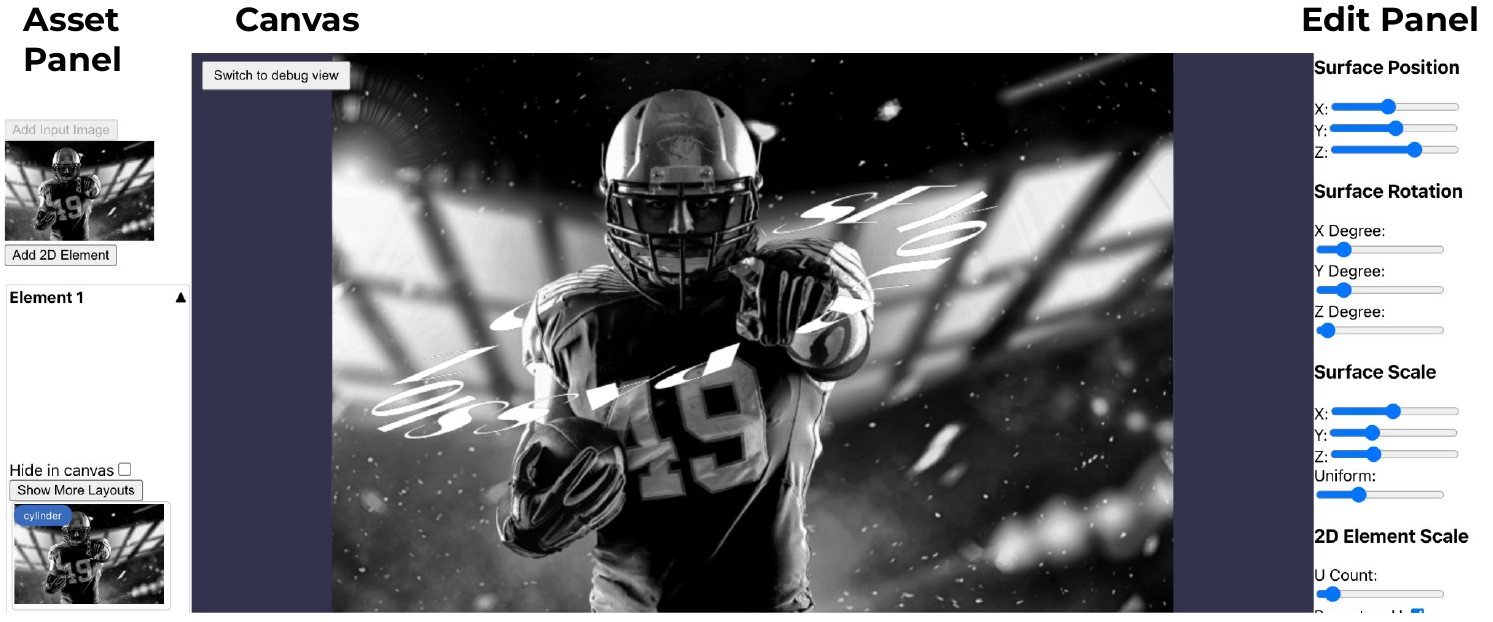}
    \caption{Proof-of-concept prototype interface.}
    \label{fig:prototype}
    \Description{A screenshot of the prototype user interface. The layout consists of three main panels: an Asset Panel on the left displaying imported assets, which is a football player; a central Canvas Edit Panel for visualizing and editing the 3D scene; and a right-side edit panel containing sliders.}
\end{figure}

We built a proof-of-concept prototype \autoref{fig:prototype} as a web interface. The prototype uses a single-image-to-3D-mesh model \textit{CRM} \cite{wang2024crm} to reconstruct the main object of a user-uploaded image into a 3D mesh. We then add primitive surfaces like planes, cylinders, and spheres to render realistic occlusion and perspective effects. Users can edit the placement of these primitives with a series of sliders that adjust the position, rotation, scaling, and content placement of the primitive surfaces. We also explored AI recommendation of 3D placement by leveraging \textit{CLIP} \cite{radford2021learningtransferablevisualmodels} and cosine similarity to retrieve 3D placement parameters based on the similarity of input images. The recommended parameters are rendered as thumbnail images for users to browse and select. More details about this prototype system can be found in our earlier paper \cite{10.1145/3706599.3719727}.

\subsection{User Study with the Prototype}
To test the feasibility, usability, and helpfulness of our proposed 2.5D creation pipeline, we conducted a user study with the prototype among nine participants aged 20-60. To understand the design challenges of both expert and novice designers, we recruited four design enthusiasts and five professional designers, all with varying levels of 2D visual design and 3D design expertise. For more details about their occupation and self-reported skill levels of 2D and 3D design, please check \autoref{tab:demo1}.

\begin{figure}
    \centering
    \includegraphics[width=1\linewidth]{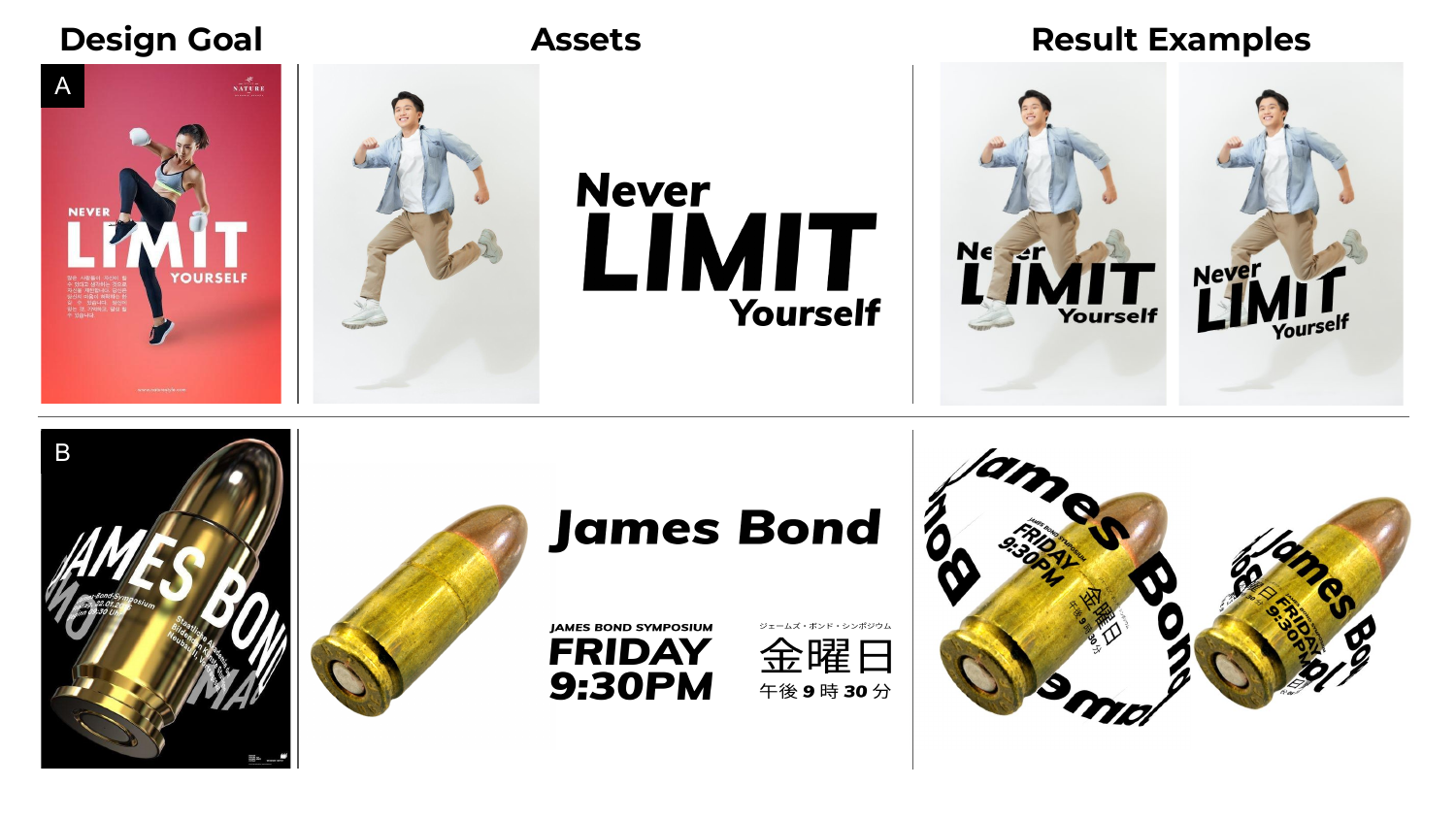}
    \caption{In our formative user study, we request participants to replicate two designs (left) with provided assets (middle) and our prototype system. Example results are shown on the right.}
    \label{fig:replication}
    \Description{A figure with two design replication tasks from a formative user study. Each task includes a design goal on the left, assets in the middle, and result examples on the right.
Row A shows a poster with a woman boxing and the phrase "Never LIMIT Yourself" as the goal. Assets include a photo of a man jumping and the same phrase in black text. The results show variations with the jumping man and phrase arranged similarly to the goal.
Row B features a bullet-shaped poster with the text "JAMES BOND" wrapped around it. Assets include a bullet image and text elements in English and Japanese. Results show bullets with wrapped text mimicking the design goal.}
\end{figure}

The study started with a brief discussion about the design genre of the 2.5D designs and also the expected workflow to create them. We then presented the prototype interface and guided participants to recreate two example designs with provided image assets (\autoref{fig:replication}). We also encouraged participants to explore the system freely and create designs as they wished. We finished the study with usability questions and Likert-scale ratings.

\begin{table*}[]
\caption{Pilot user study demographics.}
\label{tab:demo1}
\begin{tabular}{lllllrr}
\hline
Participant No. & Age              & Gender & Job Title             & \begin{tabular}[c]{@{}l@{}}Has professional \\ design experience\end{tabular} & \multicolumn{1}{l}{\begin{tabular}[c]{@{}l@{}}2D design \\ skill (1-7)\end{tabular}} & \multicolumn{1}{l}{\begin{tabular}[c]{@{}l@{}}3D design\\  skill (1-7)\end{tabular}} \\ \hline
P1              & 25-34            & M      & Product Manager       & Yes                                                                           & 5                                                                                    & 3                                                                                    \\
P2              & 18-24            & M      & Student               & No                                                                            & 5                                                                                    & 5                                                                                    \\
P3              & \textgreater{}54 & M      & Marketing Manager     & Yes                                                                           & 6                                                                                    & 3                                                                                    \\
P4              & 25-34            & M      & PhD Student           & No                                                                            & 3                                                                                    & 6                                                                                    \\
P5              & 25-34            & M      & Senior Web Operations & Yes, 2 years.                                                                 & 6                                                                                    & 1                                                                \\
P6              & 25-34            & F      & Graphic designer      & Yes, for 10 years.                                                            & 6                                                                                    & 2                                                                                    \\
P7              & 45-54            & M      & Motion designer       & Yes, for10-15 years                                                           & 7                                                                & 7                                                               \\
P8              & 25-34            & M      & PhD Student           & Not professionally                                                            & 4                                                                                    & 2                                                                                    \\
P9              & 25-34            & M      & Engineer              & No                                                                            & 5                                                                                    & 3                                                                                    \\ \hline
\end{tabular}
\end{table*}

\begin{figure}
    \centering
    \includegraphics[width=1\linewidth]{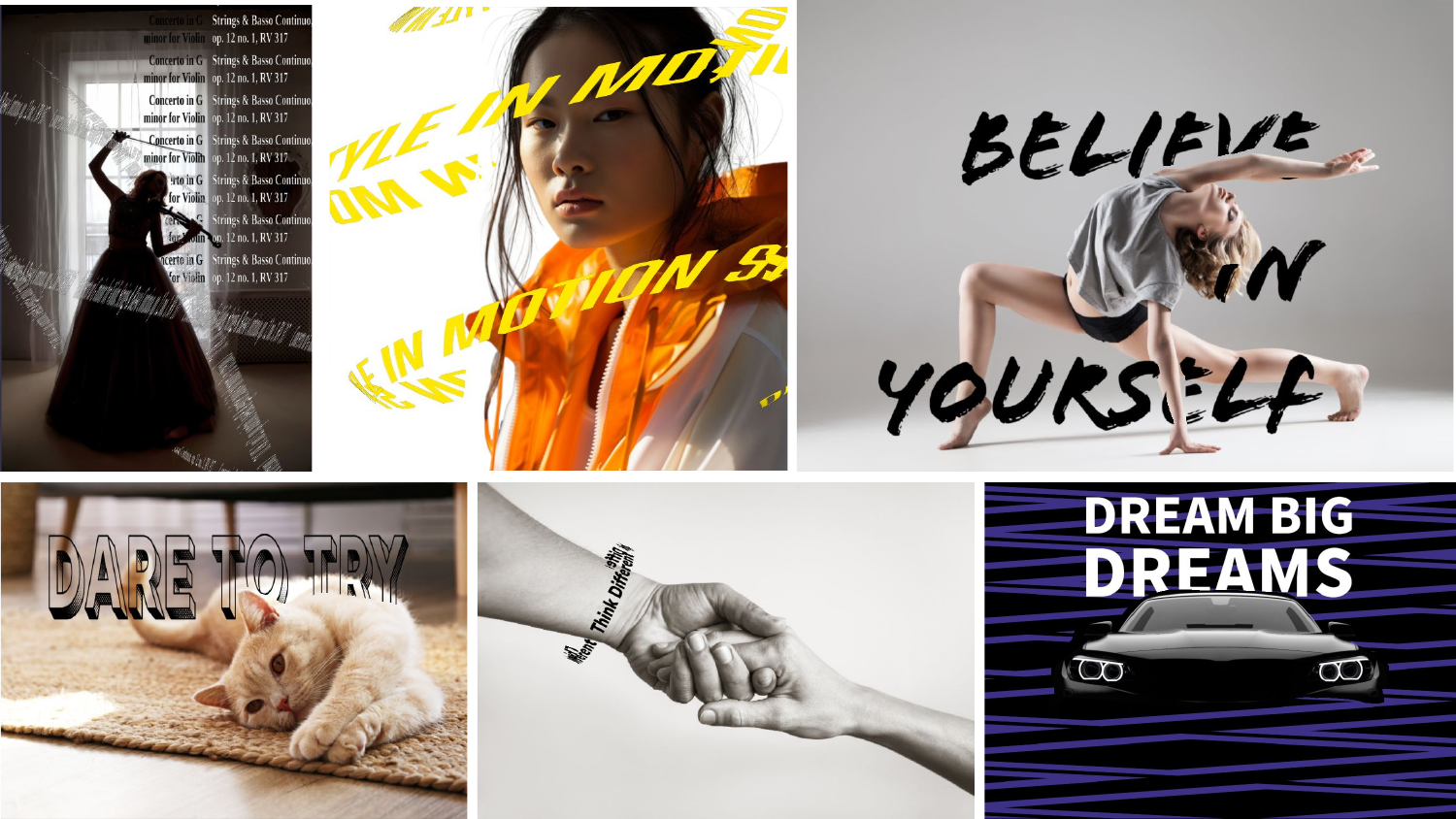}
    \caption{Design results of the open-ended exploration of the formative user study.}
    \label{fig:open-ended}
    \Description{A collage of six creative design results from participants in an open-ended user study. Designs combine photos with expressive text placement and styling.
Top left: A silhouetted woman in a gown with concert text projected across the scene.
Top center: A woman in an orange jacket surrounded by yellow, warped “STYLE IN MOTION” text wrapping around her.
Top right: A dancer in a stretching pose with the phrase “BELIEVE IN YOURSELF” in bold, expressive font.
Bottom left: A cat lying down with the phrase “DARE TO TRY” in bold 3D-style text.
Bottom center: Two hands clasped with “Think Differently” text wrapped around a wrist like a bracelet.
Bottom right: A black sports car with the phrase “DREAM BIG DREAMS” above, set against a wavy purple-striped background.}
\end{figure}

All participants successfully utilized the prototype interface to create multiple 2.5D designs, including the required replication designs and open-ended free designs \autoref{fig:open-ended}. Participants highly agree that \textit{"the prototype supported my creativity."} (avg = 4.67/5). They also commonly agree that the prototype made occlusion (avg = 4.56/5) and surrounding effects (avg = 4.56/5) easy to achieve. We observed diverse and visually compelling designs in the open-ended designs \autoref{fig:open-ended}, including designs with simple depth cues that achieve realistic partial occlusions due to accurate depth reconstruction, and also more complex designs leveraging repetition and surrounding effects. Some designs achieved serendipitous effects, like a helix of text and a purple haze of rectangles, which are created beyond expectation when exploring the parameter sliders. 

The user study reveals consistent application scenarios centered around rapid design prototyping and ideation for 2.5D visual effects. Participants consistently identified the tool as ideal for \textit{``trying out ideas at early stage''} where they could \textit{``get a few AI brainstorm suggestions''} and create \textit{``rough mock ups''} before transitioning to higher-fidelity tools (P8). Users emphasized it makes \textit{``occlusion and surround effects easier and faster''}, that would otherwise require complex 3D software knowledge (P1). The overarching theme positions DepthScape as a specialized creative sandbox for exploring pseudo-3D design possibilities before committing to production-ready tools like Photoshop or 3D software.

Importantly, the tool was perceived as valuable to both amateurs and professionals, though in distinct ways. Amateur designers valued its accessibility and freedom to experiment, noting that it \textit{``enables you to do a lot of different stuff'' }(P9). The intuitive depth cues and playful exploration encouraged creativity without requiring specialized 3D knowledge. Professionals, meanwhile, evaluated the tool through the lens of established workflows and quality standards. They appreciated the unique affordances but desired tighter integration with existing tools: as P6 remarked, \textit{``(It's okay that) We don’t have all those text editing capabilities because every other tool already has that. We don’t want to rebuild the wheel.''}. In summary, amateurs embraced it as an accessible entry point into 2.5D design, while professionals positioned it as a valuable complement to their advanced toolchains.

From these design explorations, we confirm the usability of the proposed editing method for both amateur and professional designers, and observe three ways that the prototype supports creativity: (1) the accurate depth reconstruction simplifies the creation of realistic occlusion effects; (2) the capability of rendering complex deformation and repetition effects in 3D space supports the creation of complex but orderly perspective effects; (3) the placement parameters create a design space that is big enough to contain serendipitous designs.  

\subsection{Improvements Based on Study}
Besides the overall positive findings, we also observed aspects that need improvement. We elaborate on the key findings that support our iteration of system design.

\textbf{Orthographic to Perspective.}
Due to the reconstruction feature of CRM \cite{wang2024crm} and most existing depth estimation models \cite{yang2024depth,yang2024depthV2,Patni2024ECoDepthEC,Zhu2024HABinsHA,xu2024grm,xu2024instantmesh,zhao2024flexidreamer}, the reconstruction results match the input images only in orthographic rendering. Thus, our initial prototype renders the design canvas with an orthographic camera. However, since the orthographic rendering eliminates perspective foreshortening, many participants find that the depth effects are hard to perceive and less visually impactful. In this case, we aim to switch to perspective rendering for both the depth reconstruction and the added visual elements.

\textbf{Sliders to Direct Manipulation.}
Many participants, especially those with less 3D design skills, find it hard to place elements into 3D spaces with parameter sliders. On one hand, they don't understand the meaning of the parameters and lack expectation of editing results, thus they are trapped in trial and error. On the other hand, the mapping between a conceptual goal of 3D placement, \textit{i.e.} surrounding the bullet with a ring of text (\autoref{fig:replication}B), involves a complex interplay of multiple parameters and is hard to achieve. As participants strongly prefer direct manipulation controls that match design intentions and image semantics, we aim to better parse the input images and depth reconstructions to provide simpler and smarter controls. For example, when the image contains a bullet shape, our system should be able to suggest surrounding placements that align with the bullet's direction and enable users to direct-manipulate key parameters, like the cylinder radius.

\textbf{AI Recommendation}
Our initial exploration of AI recommendations of parameters was not helpful enough (rating avg= 3.67/5) since the retrieval-based parameter matching is not precise about the image semantics and depth scene geometry. As visual designs are very specific and precise about layout, slight mismatches between image contents and 3D placements can break the alignment between parameters and visual effects. In this case, we aim to provide a more semantic-aware parsing of the input image with VLM and directly extract geometry from the depth reconstruction to ensure the precision of placement.

\textbf{Reconstructing Main Object vs Entire Scene}
The depth reconstruction model in our prototype system only captures the main object of input images, limiting the application to single-object images. More complex scenes, like urban, natural, or even a dinner table top, cannot be supported. To expand the application to more generic images, we aim to switch to models that reconstruct the depth of entire 3D scenes in the input images.

\section{The DepthScape System}
Building on the above iteration, we introduce \textit{DepthScape}, a novel Human-AI collaborative authoring pipeline that facilitates 2.5D graphic designs with depth estimation, semantic understanding, and geometry extraction. DepthScape employs a monocular geometry estimation model MoGE \cite{wang2024moge} to reconstruct 3D meshes that match visual depth cues in input images. This creates 3D spaces that can accommodate 2D design elements like text and images, and leads to realistic occlusion and perspective effects. The system further accelerates 2.5D design by parsing input images with VLM to extract key visual components and suggest parametric anchors in the 3D space that enable direct manipulative content placement.

\subsection{Reconstruct Depth Space}
\begin{figure}
    \centering
    \includegraphics[width=1\linewidth]{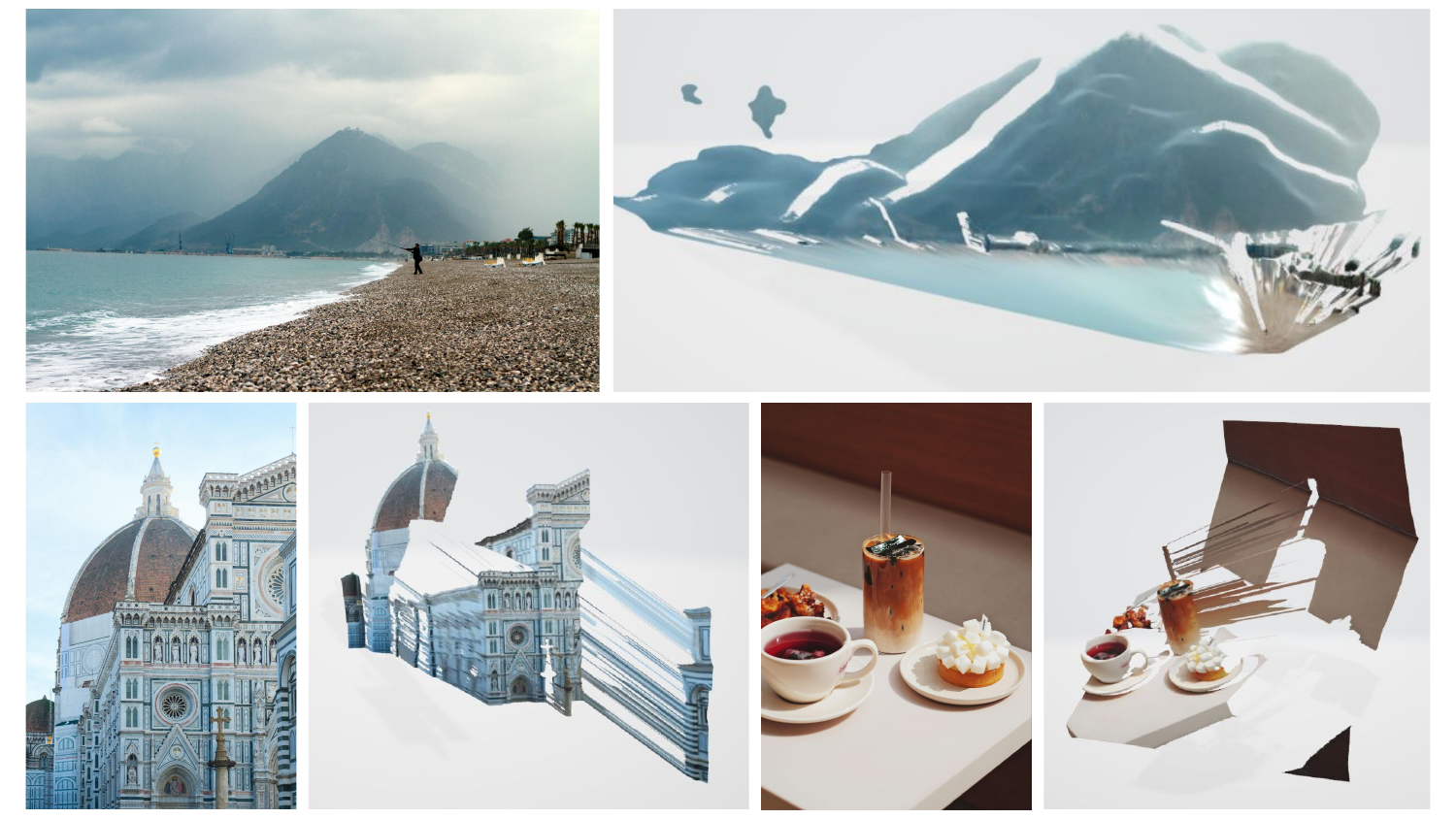}
    \caption{Examples of MoGE's geometry reconstruction based on a single input image.}
    \label{fig:MoGE}
    \Description{Figure showing examples of MoGE’s geometry reconstruction from a single input image. Each example includes an original image on the left and its corresponding 3D reconstructed geometry on the right. Top row: a beach scene with mountains is reconstructed into a 3D terrain. Bottom row (left to right): a historic cathedral with domes is reconstructed into a multi-faceted 3D structure; a table with drinks and desserts is reconstructed into a 3D scene with extruded table surfaces and objects.}
\end{figure}

We use \textit{MoGE} \cite{wang2024moge} to create a 3D mesh that reflects visual depth cues in a given 2D image. Unlike many existing depth estimation models, MoGE conducts affine-invariant monocular geometry estimation, with reconstruction results that match the input images in perspective rendering (\autoref{fig:MoGE}). This reconstructed 3D mesh encodes critical depth cues and serves as the foundation of our creation pipeline.

\begin{figure*}
    \centering
    \includegraphics[width=0.8\linewidth]{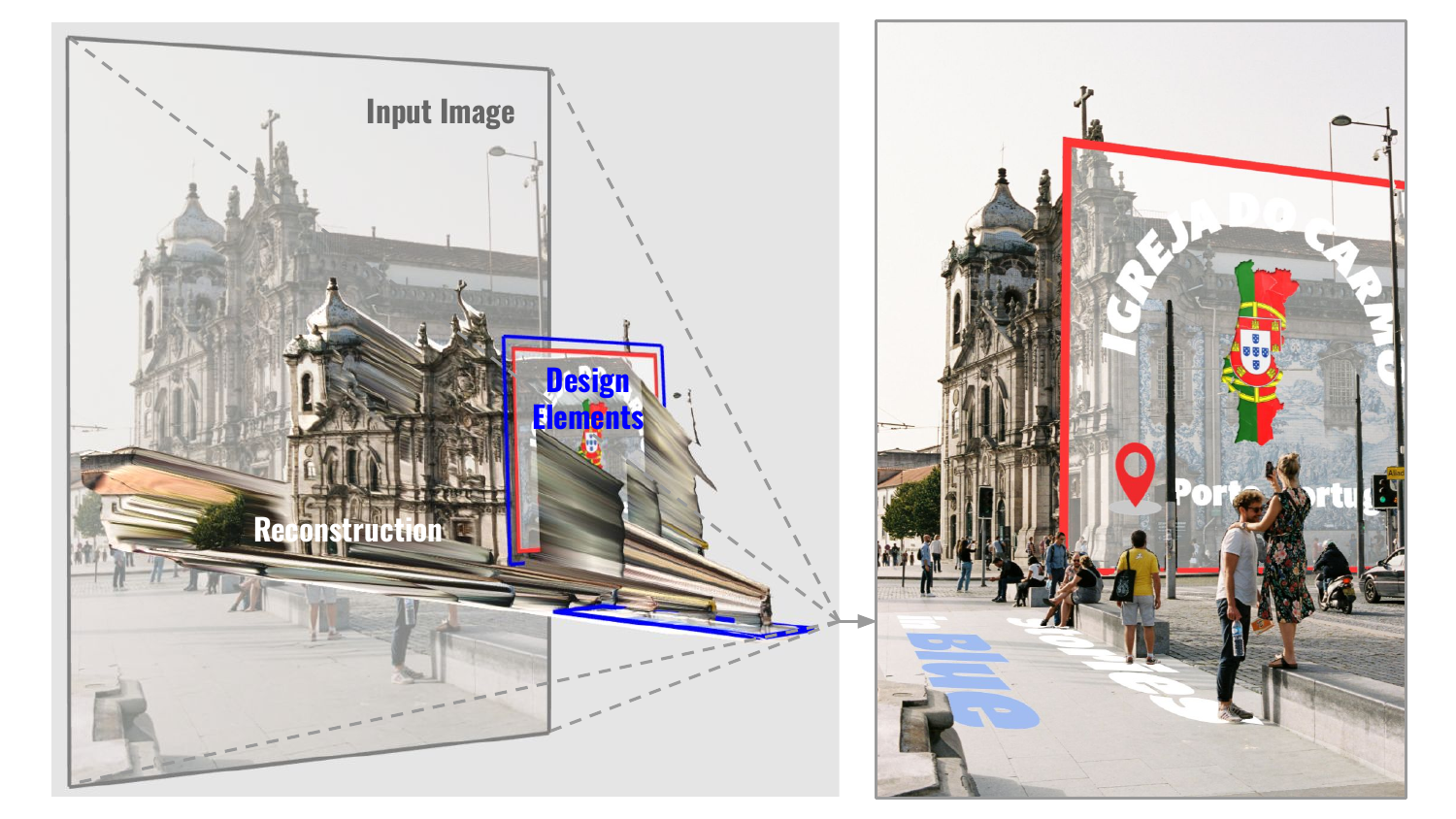}
    \caption{DepthScape's 3D space composition and and rendering results. Left: 3D space viewed from a different angle. Text shows the original input image, the reconstructed pointcloud, and also the added in design elements. Gray dash indicates the camera position. Right: render results with design elements embedded into the scene of the input image.}
    \label{fig:render}
    \Description{Figure showing DepthScape’s 3D space composition and rendering results. The left side illustrates the process: an input image of a historic church facade is reconstructed into a 3D space with layers extending backward and overlaid with labeled “Design Elements.” The right side shows the final rendered result with virtual design elements (e.g., map pin, text, and a Portuguese flag) integrated seamlessly onto the 3D scene, enhancing the visual experience of the original photo.}
\end{figure*}

We place the 2D visual elements and the reconstructed depth mesh into a shared 3D space to render realistic occlusions and natural perspective effects (\autoref{fig:render} left). By default, rendering this 3D scene correctly produces occlusions between the depth mesh and the added 2D content. However, it also exposes the reconstructed 3D mesh itself, whose visual quality is noticeably lower than the original image. To address this, we engineer the rendering pipeline so that the 3D mesh remains fully functional for occlusion but is not visible. Specifically, we disable color pixel writing for the depth mesh, effectively making it invisible while preserving its occlusion behavior. This creates a “ghost object” that blocks other elements, hence leaving holes in the rendered image. We then fill these holes using pixels from the original input image: the raw image is rendered once as a background layer in an earlier render pass. As a result, the final output appears as if objects in the original image are naturally occluding the newly added visual elements (\autoref{fig:render} right—note how the human figures occlude the text layers).

\subsection{Simplify 2.5D Design with Parametric Anchors}
During our formative user study, we observed challenges in precisely placing design elements in reconstructed 3D spaces. However, we also identified two design patterns that simplify the 3D placement task: (1) Design elements are often anchored to objects in the input images, \textit{e.g.}, wrapping text around a human arm or aligning a plane with a building facade. (2) Designers explore variations by adjusting one or two key parameters while keeping others fixed to preserve design semantics, \textit{e.g.}, shifting a plane along its normal vector while keeping it parallel to a building.

\begin{figure}
    \centering
    \includegraphics[width=1\linewidth]{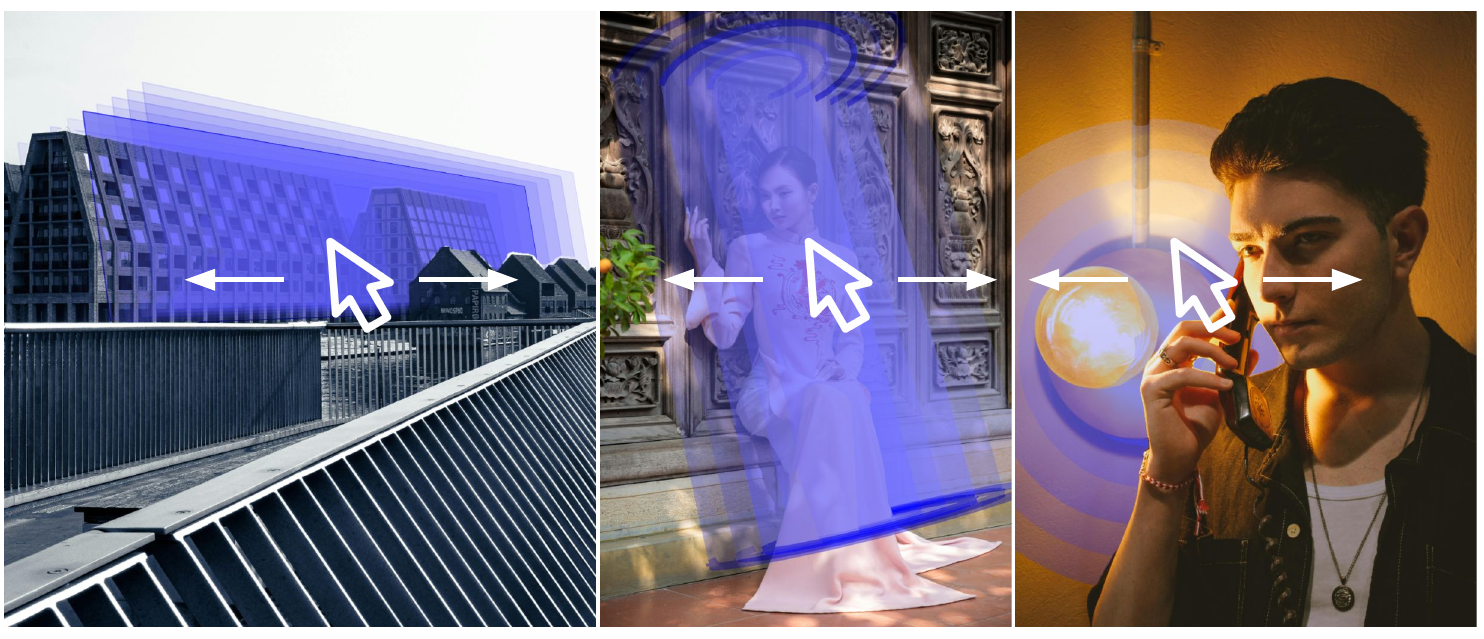}
    \caption{DepthScape currently supports three types of parametric anchors that can be direct manipulated by mouse moving. From left to right: Planar, Cylindrical, Spherical. Note that the anchor in the spherical example is centered on a glass sphere structure that is partially occluded by the blue contour.}
    \label{fig:direct-manipulation}
    \Description{Figure showing DepthScape’s support for three types of parametric anchors that can be manipulated with mouse movement. From left to right: (1) a planar anchor applied to a modern building facade with stacked blue translucent planes; (2) a cylindrical anchor wrapping a blue translucent layer around a figure in front of a decorative door; (3) a spherical anchor surrounding a light bulb, beside a person using a telephone, shown with concentric circular overlays. Each example includes a white cursor and double-headed arrow indicating mouse following interaction.}
\end{figure}

To formalize these patterns, we borrow the concept of \textit{anchors} from augmented reality research, to create \textit{parametric anchors} that binds visual contents to objects in images. Specifically, parametric anchors are focused on a certain object in the input image, centering and/or aligning the content placement to the object's geometric properties. For example, planes in parallel to a building facade, cylinders surrounding a human figure, and spheres that center around a glass bulb (\autoref{fig:direct-manipulation}). We use parametric anchors to constrain placement parameters and reduce degrees of freedom to the most relevant ones. For example, in a Planar parametric anchor, all placements share a fixed reference plane and can only shift along a predefined axis (\textit{i.e.}, the normal vector). Similarly, a Cylindrical parametric anchor organizes placements along a shared central axis, allowing only radius or height adjustments, while a Spherical parametric anchor centers placements around a fixed point, varying only in radius or angular position.

By leveraging these parametric anchors, we transform unconstrained 3D placements into structured adjustments that align with 3D scene geometry. This enables direct manipulation in a 2D viewport, which would otherwise be ambiguous. For instance, translating a plane freely in 3D from a 2D perspective is ill-defined, but within a Planar parametric anchor, it reduces to a simple 1D adjustment along the normal vector, which can be directly mapped to mouse input. See \autoref{fig:direct-manipulation} for examples.

\subsection{Extract Parametric Anchors using Code Synthesis}
 To automatically extract such parametric anchors from input images, we leverage a VLM (gpt4o \cite{hurst2024gpt}) to semantically parse the input image and guide geometry extraction processes. Inspired by prior work \cite{gupta2023visual}, we prompt VLM to follow a visual program template and enable modular and sequential geometry extraction that focuses on certain visual components of the input images. Each VLM-generated visual program will be based on a certain design intention, and expands into a sequence of program cells. Such a sequence includes three main parts: segment the input image, extract 3D geometry, create parametric anchors. Each part has multiple options of visual coding cells, and the combination creates a variety of extraction abilities.

\textbf{Segment the input image}. A visual coding cell \textit{Text2Mask (prompt)} can segment the original input image with a text prompt, which is generated in place along with the visual program. We parse the text prompt and use grounded-SAM-2 \cite{ravi2024sam2segmentimages} to generate corresponding image masks. Subsequently, we select the corresponding point clouds as vertices of the reconstructed depth mesh, with a \textit{Mask2Pointcloud (mask)} cell, which performs pixel-wise selection of the points that fall within the mask. We then use these selected point clouds for further geometry extraction.

\textbf{Extract 3D geometry}. Depending on the image semantics, we extract a variety of geometries from the segmented point clouds. 

\textit{Pointcloud2Plane (pointcloud)}: For planar objects like building facades, grounds, water surfaces, and table tops. Implemented with PyRANSAC-3D \cite{mariga_pyRANSAC3D} to extract planes from the point clouds. We also implement several derived geometries from an extracted plane, like the primary direction  and the perpendicular extrusion, to enrich the results and empower more complex geometry references.

\textit{Pointcloud2Cylinder (pointcloud, direction)}: For elongated objects like human figures, tree trunks, or trains. This cell extracts a containing cylinder, with an optional parameter of axis direction. When direction is not provided, we use PCA to find the primary component of the point cloud as the axis direction. We then find the center location that contains all the points with the smallest radius. 

\textit{Pointcloud2Sphere (pointcloud)}: For rounded shaped objects. This cell extracts a center and a minimum radius to contain the point cloud. 

Besides these cells that extract primitive surfaces, we also implement cells that extract human skeleton and facial structure due to the high frequency of human figures in both our formative study and our wider online search. 

\textit{Pointcloud2Skeleton (mask)}: This cell extracts key pose landmarks in the masked region with mediapipe \cite{lugaresi2019mediapipe} and casts the 2D coordinates to the depth point cloud to yield 3D skeleton points. This extracted 3D skeleton can then yield frontal and median planes (as \textit{SKELETON.frontal} and \textit{SKELETON.median} in our visual programs) to support content placement. We also implement derived directions, like upward direction as \textit{SKELEON.cranial}, and front direction as \textit{SKELEON.anterior}, from skeletons.

\textit{Pointcloud2Face (mask)}: For human figures zooming in more to faces, we extract the face geometries. Similar to skeleton extraction, this cell extracts key facial landmarks with mediapipe \cite{lugaresi2019mediapipe} and casts the 2D points to depth point cloud to yield 3D face points. This extracted 3D face can then yield frontal and median planes (as \textit{FACE.frontal} and \textit{FACE.median} to support content placement. Derived directions like upward direction as \textit{FACE.cranial}, and front direction as \textit{FACE.anterior} are also supported.

\textbf{Create parametric anchors}. We then wrap extracted geometries into parametric anchors and finish the visual coding sequence. 

\textit{Planar(plane)}: Create a planar parametric anchor with an extracted plane or an derived plane. 

\textit{Cylindrical(cylinder)}: Create a cylindrical parametric anchor with an extracted cylinder. 

\textit{Spherical(sphere)}: Create a cylindrical parametric anchor with an extracted sphere.

\begin{figure}
    \centering
    \includegraphics[width=0.6\linewidth]{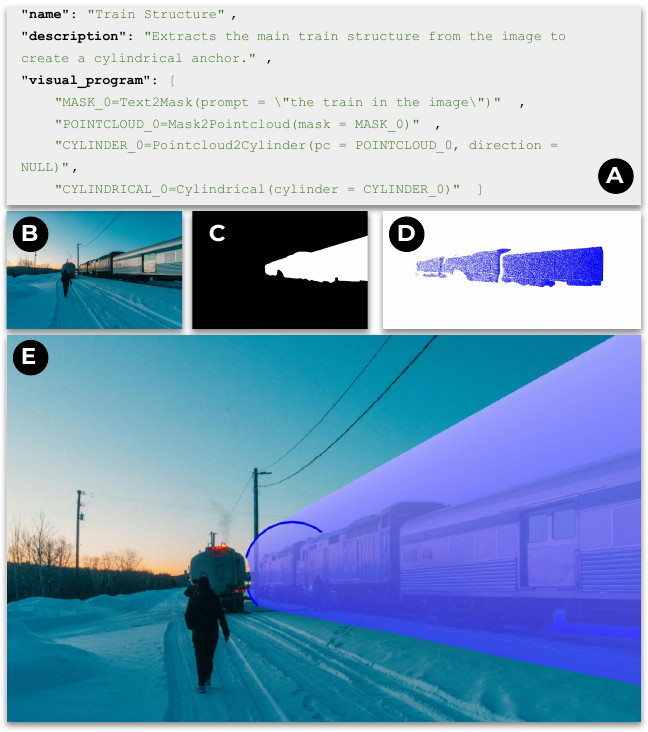}
    \caption{DepthScape employ a vision language model to (A) generate visual programs based on (B) input images. The program guides a geometry extraction process, including (C) masking a certain image part, and (D) creating the point cloud. The final result would be (E) interactive parametric anchors aligned with objects in the scene. }16
    \label{fig:extraction}
    \Description{A multi-panel figure demonstrating the DepthScape system for extracting geometric structures from images using vision-language models. Panel (A) shows a visual program script that defines steps to extract a cylindrical anchor from an image of a train. Panel (B) displays the original input image of a person walking toward a train on a snowy track. Panel (C) shows a binary mask isolating the train. Panel (D) presents a 3D point cloud representation of the masked region. Panel (E) illustrates the final result with a blue semi-transparent cylindrical anchor overlaid on the train in the original image, representing the extracted geometry aligned with the object. The caption explains how the system converts image and language input into interactive 3D anchors.}
\end{figure}

By combining different visual coding cells, we can extract various types of geometry from depth reconstructions based on image semantics. See \autoref{fig:extraction} for an example extraction, including the visual program content, steps involved, and the final result.

We prompt VLM with a list of 16 visual program to enforce the formatting and general design patterns, each includes description of the raw input image and a matching visual program based on the content. See \autoref{appendix:examples} for examples. VLM then generates semantically reasonable visual programs based on the input images. DepthScape parses the generated visual programs and conducts the encoded extraction pipeline to create parametric anchors. The successfully extracted parametric anchors, along with the original design rationale, will be shown in our design interface for users to interact with.

\subsection{The DepthScape Interface}
\begin{figure*}
    \centering
    \includegraphics[width=1\linewidth]{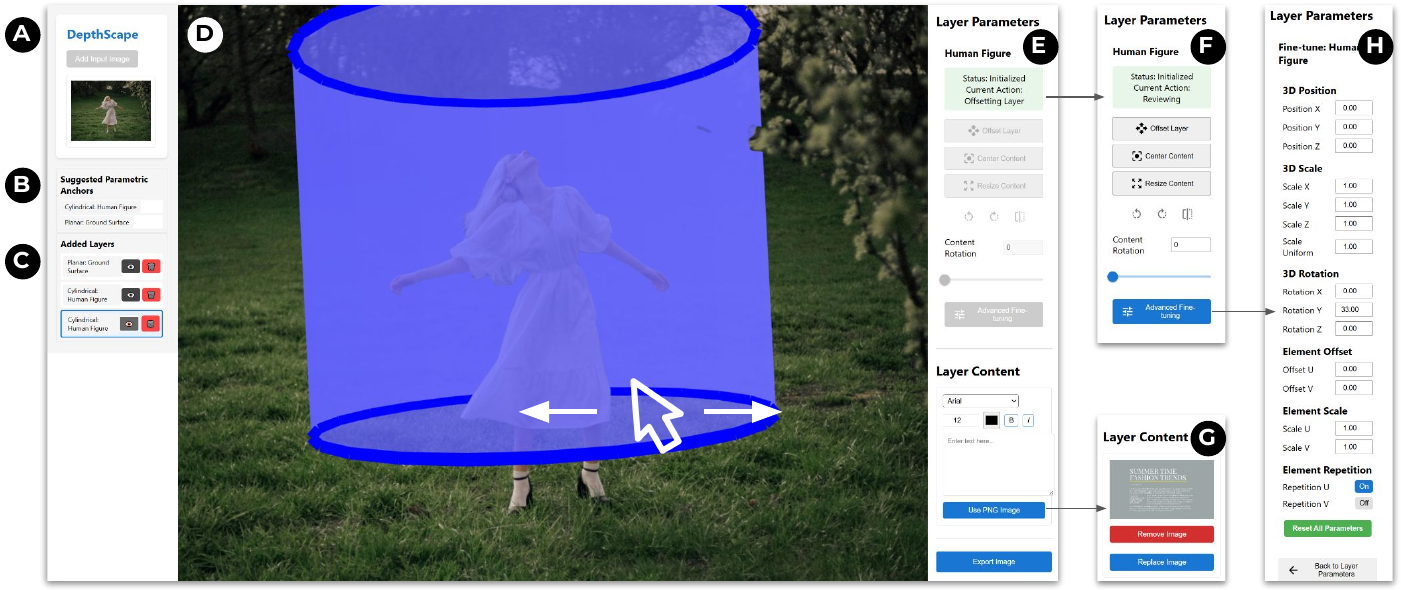}
    \caption{DepthScape's user interface. (A) The input button and the user uploaded image. (B) AI-suggested parametric anchors. (C) Added layers into the scene. User can select, hide/unhide or remove these layers. (D) The editing canvas that enables direct manipulation edits. (E) Parameter panel for the selected layer, during initialization. (F) After initialization, more editing options are enabled, including rotation of content, adding text or (G) image content, and (H) fine-tune the layer parameters.}
    \label{fig:UI}
    \Description{A user interface of DepthScape for editing and placing parametric 3D anchors. Panel (A) shows the sidebar with an image upload button and a preview of the uploaded image. Panel (B) lists AI-suggested parametric anchors, including a cylindrical human figure and planar ground surface. Panel (C) displays the currently added layers with visibility and delete toggles. Panel (D) is the main canvas showing a user editing a blue translucent cylindrical anchor over a person in the photo, with an arrow cursor indicating resizable boundaries following mouse move. Panel (E) shows the parameter panel during layer initialization. Panel (F) reveals the same panel after initialization with advanced fine-tuning options enabled. Panel (G) allows users to add or replace image content within a layer. Panel (H) provides fine-tuning parameters for 3D positioning, scaling, rotation, offset, and repetition settings. The figure illustrates interactive editing of layered geometric content.}
\end{figure*}

The DepthScape interface is implemented with \textit{React} and \textit{Babylon.js} as a webpage (\autoref{fig:UI}). The interface has three main parts: (1) the Asset Panel imports input images and shows suggested parametric anchors, as well as added layers in the current scene; (2) the canvas renders the 2.5D effects and enables direct manipulation of design elements; (3) the editing panel lists editing options for selected parametric anchors, including adding text or image contents, and fine-tuning the design with parameters.

The user first imports an input image (\autoref{fig:UI}A), which is processed to reconstruct the 3D depth mesh and extract suggested parametric anchors in real time. Then the user can inspect the suggested parametric anchors by hovering on the enlisted options (\autoref{fig:UI}B), and click to select ideal ones.This triggers an initialization sequence for the selected anchor, which includes offsetting the surface, resizing the surface (for planar parametric anchors), centering the content’s UV coordinates on the anchor, and scaling the content UVs. Each step is presented as a mouse-following direct-manipulation operation that the user can confirm with a left-click or skip with a right-click. After finishing the initialization sequence, the user can rotate and mirror the 2D content with buttons and a slider (\autoref{fig:UI}F, lower half). To further fine-tune the placement, we also included an advanced fine-tune panel which enables users to adjust all 3D parameters of the placement (\autoref{fig:UI}H). When initially creating the placement, a white ``LOREM IPSUM'' text will be shown as placeholder content to help the user conduct placement. The user can replace the placeholder by typing in text or uploading images (\autoref{fig:UI}G). The final results can be exported as a rendered image with one click. 

\section{Evaluation}
We evaluate DepthScape's performance in terms of efficiency, robustness, output diversity, and design quality. We construct a diverse image collection and run it through the DepthScape system to inspect the processing time, error rate, distribution of AI-suggested parametric anchors, as well as the design quality.

\subsection{Collecting Test Images}
\begin{figure}
    \centering
    \includegraphics[width=1\linewidth]{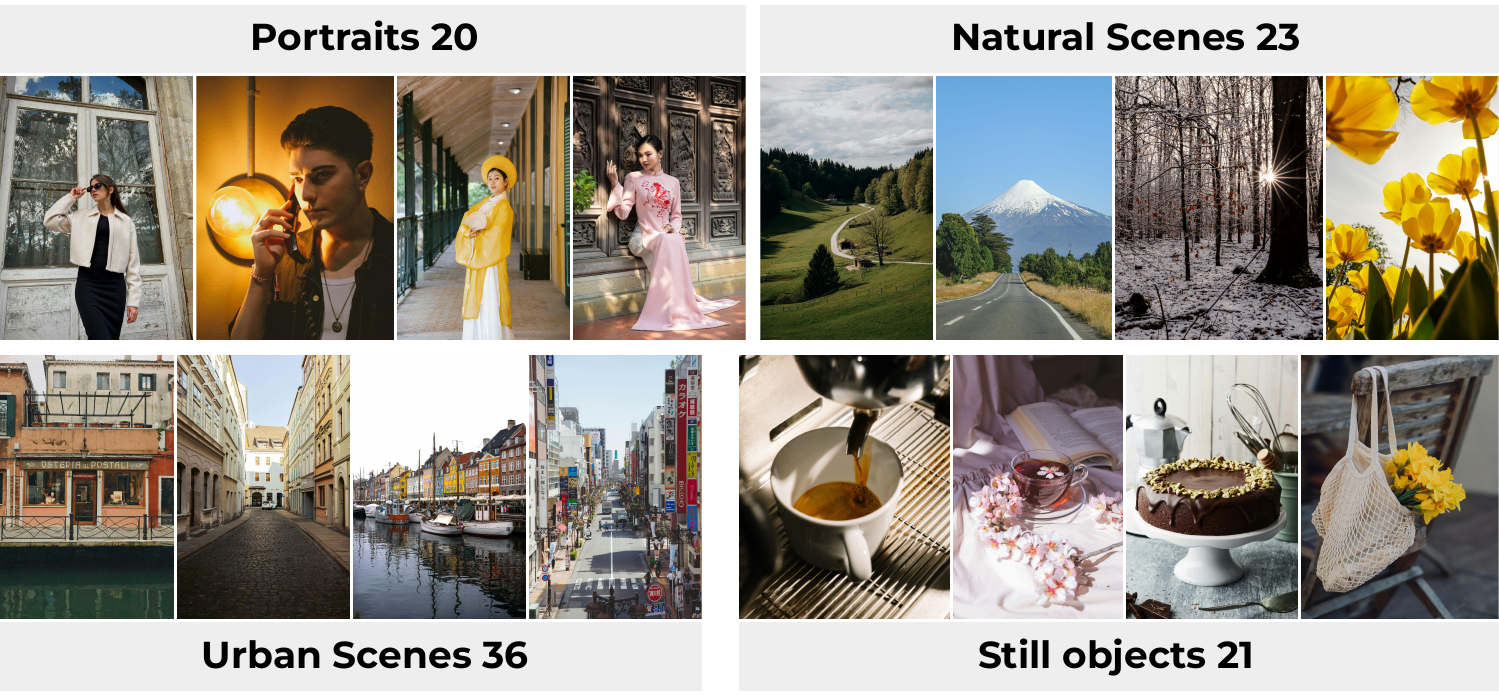}
    \caption{Four types of test images.}
    \label{fig: test set}
    \Description{A collage image showing four types of images in the test set. There are 20 portrait images shown on top left; 23 natural scenes shown on top right; 36 urban scenes shown on bottom left; 21 still objects shown on bottom right.}
\end{figure}

To ensure a diverse and non-biased selection of test images that reflects real-world designer needs, we collect images from Pexels \cite{pexels}, which is a free-stock images sharing platform. We download from Pexels' waterfall of trending images, which are constantly updated to reflect the popular choices of the designer community. We collected 50 images on two separate dates, creating a pool of 100 high-quality and diverse images. We observe four main content types from this collection: portrait, urban/building, nature/animal, and static objects. \autoref{fig: test set} shows the distribution and example images.

\subsection{Procedure}
We sequentially upload all 100 collected images to DepthScape, and inspect the processing results including the processing time, visual programs generated by VLM, extraction results of each visual program, and also error logs of failure cases. We analyze the log data to understand the efficiency and robustness of the DepthScape system, and also the distribution of parametric anchor types and causes of errors. 

\begin{figure}
    \centering
    \includegraphics[width=1\linewidth]{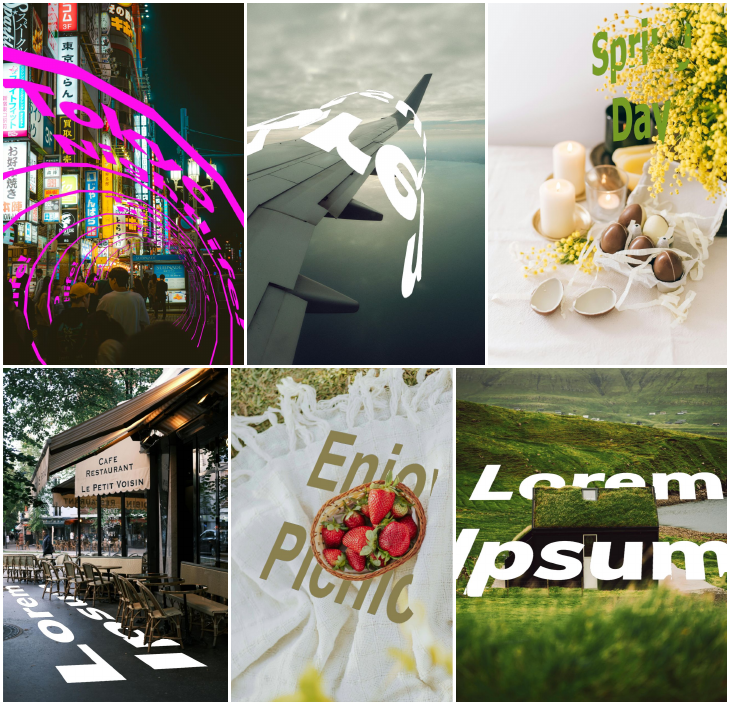}
    \caption{Example design results shown to experts.}
    \label{fig:examples}
\end{figure}

Additionally, we evaluate the quality of the suggested parametric anchors and final outputs with an expert review session among three professional designers. See \autoref{tab:demographics} for their demographics information, including their years of experience and design fields. We showcased DepthScape's UI recording along with the recommended parametric anchors for four images, one from each type of our image collection. We also presented several design results created by the authors to better showcase the system output quality (see \autoref{fig:examples}). We then collect Likert-scale ratings of the quality of recommended parametric anchors and design outputs, while also soliciting comments regarding the application scenarios and system improvements. The UI recording and example designs shown in the expert review sessions are included in the supporting files.

\begin{table}[]
\caption{Expert review demographics}
\label{tab:demographics}
\begin{tabular}{lllrl}
\hline
\rowcolor[HTML]{FFFFFF} 
Expert & Gender & Age   & \multicolumn{1}{l}{\cellcolor[HTML]{FFFFFF}\begin{tabular}[c]{@{}l@{}}YOE as\\  Designer\end{tabular}} & Design Type                                                                \\ \hline
\rowcolor[HTML]{EFEFEF} 
E1          & Male   & 25-34 & 1.5                                                                                                    & UI/UX                                                                      \\
\rowcolor[HTML]{FFFFFF} 
E2          & Female & 25-34 & 12                                                                                                     & Game, UI/UX                                                                \\
\rowcolor[HTML]{EFEFEF} 
E3          & Male   & 45-54 & 25                                                                                                     & \begin{tabular}[c]{@{}l@{}}Web, Animation, \\ Motion Graphics\end{tabular} \\ \hline
\end{tabular}
\end{table}

\subsection{Technical Results}
97/100 of our test images successfully went through the DepthScape system to generate at least one parametric anchor based on image semantics. On average, processing of each image took 21.1 seconds (std=5.5s) and generates 4.2 (std=1.1) visual programs. 89\% of all generated visual programs were successfully processed to yield parametric anchors available for user interaction.

\begin{figure*}
    \centering
    \includegraphics[width=1\linewidth]{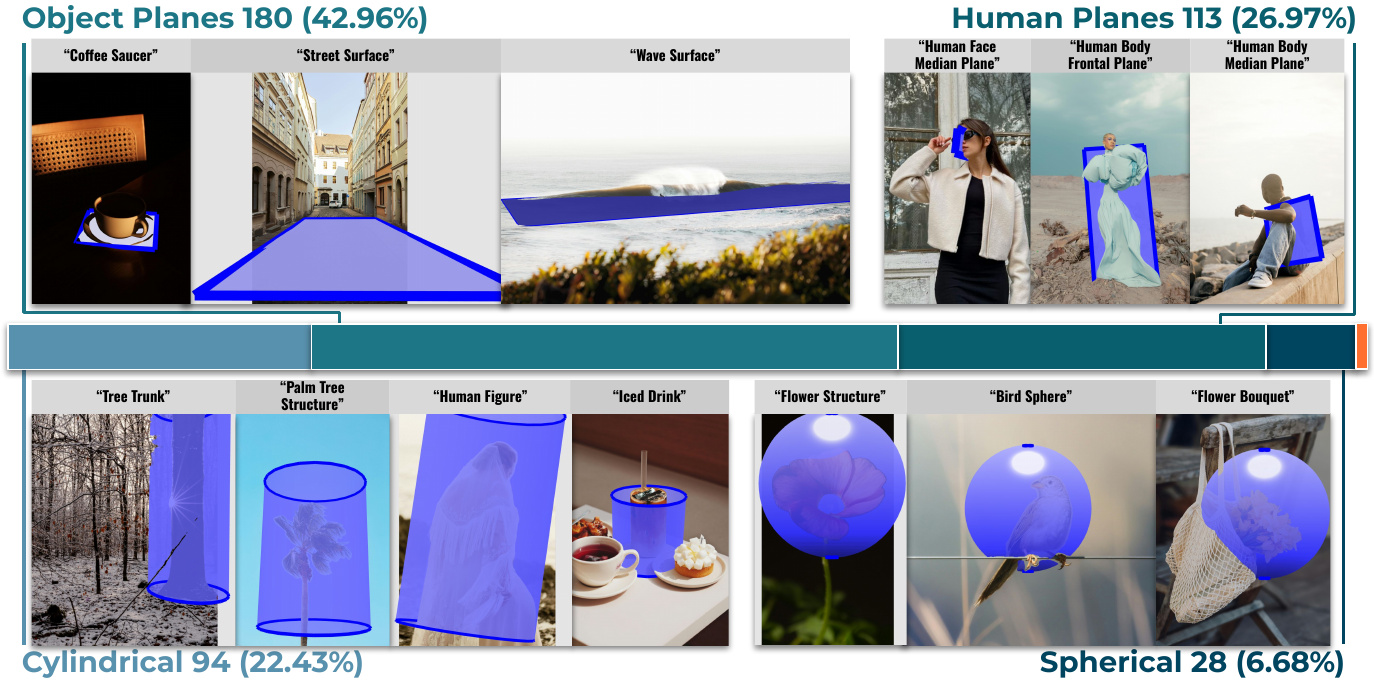}
    \caption{Four types of parametric anchors. Each example shows the name of the parametric anchor on top.}
    \label{fig:results}
    \Description{A categorized visual summary of parametric anchors used by DepthScape, divided into four types with their respective counts and percentages. At the top, "Object Planes" (180 anchors, 42.96\%) are shown with examples like “Coffee Saucer,” “Street Surface,” and “Wave Surface,” each with flat blue overlays. Next, “Human Planes” (113 anchors, 26.97\%) include overlays aligned to body parts, such as the “Human Face Median Plane” and “Human Body Frontal Plane.” Below that, “Cylindrical” anchors (94 anchors, 22.43\%) depict blue cylindrical overlays fitted around vertical structures, including “Tree Trunk,” “Palm Tree Structure,” and “Human Figure.” Lastly, “Spherical” anchors (28 anchors, 6.68\%) are shown encapsulating round objects like flowers and birds in blue spheres, with examples such as “Flower Structure” and “Bird Sphere.” The figure highlights the distribution and application of different anchor types across varied real-world images.}
\end{figure*}

\begin{figure}
    \centering
    \includegraphics[width=0.6\linewidth]{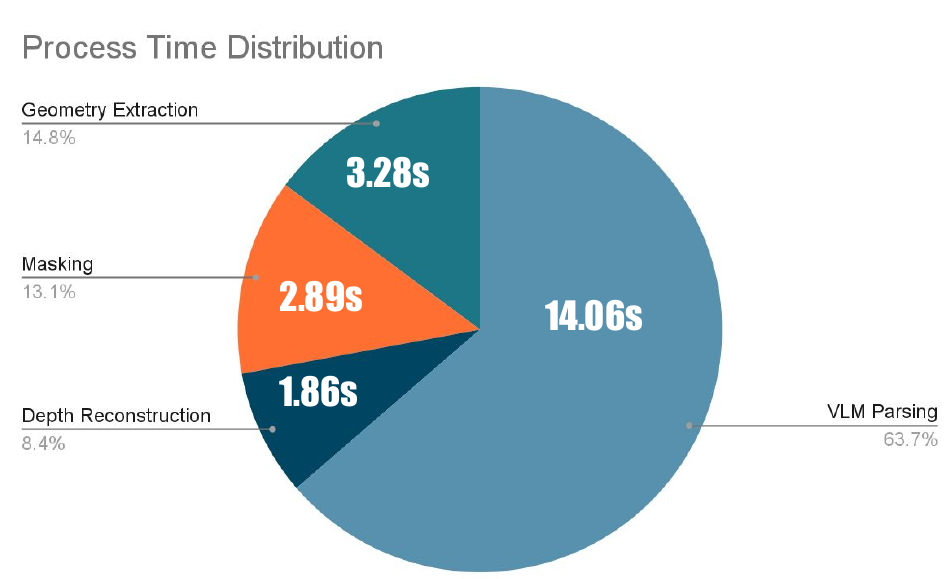}
    \caption{The distribution of processing time.}
    \label{fig:time}
    \Description{A pie chart showing the distribution of procssing time. More than half of the processing time (63.7\%, avg=14.06s, std=4.04s) is spent waiting for GPT responses, while the processing time for depth reconstruction (8.4\%, avg=1.86s, std=2.02s), masking (13.1\%, avg=2.89s, std=1.23s; each masking time avg=0.89s, std=0.19s), and geometry extraction (14.8\%, avg=3.28s, std=1.30s) takes up the remaining ~40\%.}
\end{figure}

\autoref{fig:time} shows the averaged distribution of processing time. More than half of the processing time (63.7\%, avg=14.06s, std=4.04s) is spent waiting for GPT responses. Benefiting from the progress in VLM, such time can be greatly reduced in future iterations with newer and faster models. The processing time for depth reconstruction (8.4\%, avg=1.86s, std=2.02s), masking (13.1\%, avg=2.89s, std=1.23s; each masking time avg=0.89s, std=0.19s), and geometry extraction (14.8\%, avg=3.28s, std=1.30s) takes up the remaining ~40\%.

In total, 419 visual programs were generated by VLM, among them 370 successfully went through geometry extraction processes and yielded interactive parametric anchors. \autoref{fig:results} shows the distribution of parametric anchor types of all generated visual programs. The most common parametric anchor type is Planar, which we further break down as object planes (180 occurrences, 42.96\% of all) and skeletal/facial planes (113 occurrences, 26.97\% of all). VLM also suggested 94 Cylindrical and 28 Spherical visual programs. See \autoref{fig:results} for the distribution and examples.


\begin{figure}
    \centering
    \includegraphics[width=0.8\linewidth]{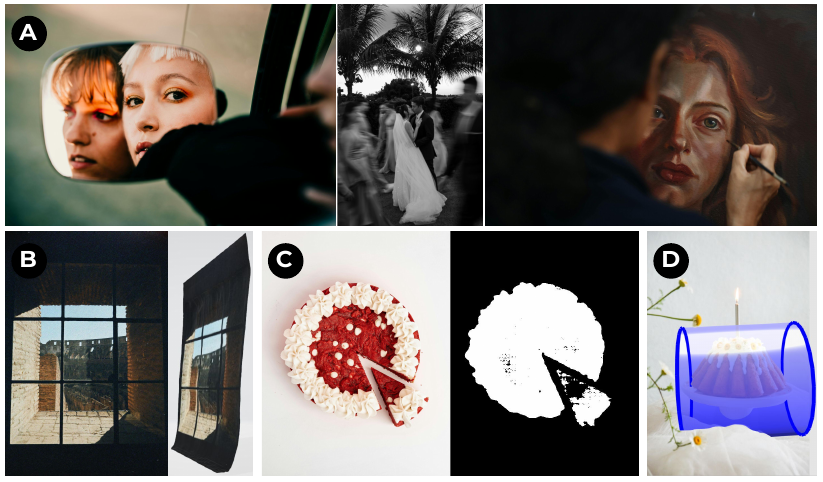}
    \caption{Failed cases. (A) Three images with no parametric anchor generated. (B) Depth reconstruction on a transparent glass surface loses the geometry behind the glass. (C) The masking of "the cake slice" ended up with the entire cake. (D) The cylinder fitting fails to capture the shorter but symmetrical axis.}
    \label{fig:fails}
    \Description{A visual summary of failure cases in DepthScape’s parametric anchor generation. Panel (A) displays three images where no anchors were created: a photo of two faces in a car mirror, a blurry black-and-white wedding scene, and a close-up of a painter working on a portrait. Panel (B) shows a depth reconstruction failure where a window reflects a scene, but the transparent glass prevents accurate geometry capture. Panel (C) highlights a masking error, where the intent was to isolate a cake slice, but the entire cake was masked instead. The original image of a red velvet cake and its binary mask are shown side by side. Panel (D) demonstrates a fitting issue where a blue cylindrical anchor is applied incorrectly to a small symmetrical cake, failing to capture its true vertical proportions. The figure illustrates common challenges in visual grounding and geometry extraction.}
\end{figure}

We observe two types of errors in the 46 failed visual programs, which are all Planar visual programs. The most common error type (34 occurrences) is failure in detecting human skeletal and facial landmarks. This is due to incomplete or failed masking of human figures. The other type (12 occurrences) was errors in visual program formatting, \textit{e.g.} hallucinative visual program cells, not ending with a valid parametric anchor, or program cells like \textit{Linear} or \textit{PLANE.median}, which is a semantically reasonable extension from the provided examples but not directly indicated to be supported. \autoref{fig:fails}A shows three failed images, whose human figure extraction failed.

Besides these hard error cases, we also observed errors that lead to unsatisfactory parametric anchors. In \autoref{fig:fails}B, the depth reconstruction of a glass surface was flat and lost all geometry behind the transparent surface. However, VLM still suggested geometry fitting for the ground surface behind the glass, leading to misaligned results. In \autoref{fig:fails}C, VLM suggested to mask \textit{``the cake slice''} to focus on the smaller slice in the image, but the segmentation model ended up masking the entire cake, leading to another misalignment case. In \autoref{fig:fails}D, the cylinder fitting module tried to capture the cake geometry, but adopted a horizontal axis direction, instead of the shorter but symmetrical and semantically valid vertical axis.

\subsection{Expert Review Results}
Experts rated our parametric anchor quality positively (7, 4, 5 on a 1-7 scale) as well as the design output quality (6, 4, 5). They appreciated DepthScape’s ability to parse complex scenes and generate diverse parametric anchors based on image content. E1 remarked, \textit{``I can totally imagine that a Vogue cover designer would use it very powerfully.''} Experts noted that the current system is already well-suited for design hobbyists. E2 highlighted its accessibility, stating, \textit{``I think the value of this is that 99\% of people who have no idea how to do this [2.5D effects] will benefit a lot from this.''} However, they also pointed out that if intended for professional use, there are several areas for improvement, \textit{e.g.} occlusion precision, rendering quality, and perspective accuracy. For instance, E3 observed that while the cylinders and spheres were effective, plane edges were sometimes slightly misaligned with the base objects, requiring manual adjustments to meet professional standards.

Design experts proposed various application scenarios. E1 suggested DepthScape could be used for creating book covers, T-shirts, and even memes. They also envisioned the possibility of generating 2.5D animations by simply binding a position parameter to time. E2 emphasized utility in design ideation, stating, \textit{``I think it's nice to use [DepthScape] to quickly put together some concepts. It's definitely helpful to accelerate the workflow.''} They also noted its potential for users of platforms like Canva or Instagram who want to create striking text effects without needing expertise in complex software like Photoshop. E3 suggested additional use cases, including UI mockups, presentation landing pages, and promotional materials. E3 was particularly interested in DepthScape’s ability to analyze images in terms of both semantics and depth, believing that its capacity to break apart images could support the creation of video effects in tools like After Effects.

Experts also expressed a desire for additional features to enhance DepthScape’s functionality. E3 hoped for more sophisticated content anchors that incorporate multiple directional cues from the scene. They also emphasized the need for text readability enhancements when adding text elements. E2 envisioned DepthScape being integrated into existing design platforms like Figma, Canva, or Instagram to streamline content creation. All three experts requested improved blending between the depth scene and added content, including support for light estimation, shadow casting, and blur effects to further enhance realism. Additionally, E2 and E3 advocated for the ability to export designs in editable, multi-layer formats (\textit{e.g.}, PSD files) for further refinement in professional design software.

\section{Application Scenarios}
Inspired by our formative study, expert review, as well as our exploration, we discuss five application scenarios to further demonstrate the creative power and future potential of DepthScape. Please also check the video figure for additional details and animations.

\subsection{Integration into Image Editing Tools}
Suggested by both our formative user study and expert review, the DepthScape system can work as a good complementary tool for existing professional image editing tools.
If integrated, the depth-based editing in these image editing pipelines can be accelerated, amateur users who do not know how to create these effects can be encouraged and enabled, and also the final output quality can benefit from other powerful editing features. In order to achieve this, one way is to include DepthScape as a plug-in for these platforms. Another simpler possibility is enabling DepthScape to export design results as layered and editable files 
and import them into existing platforms for further editing.

\subsection{Video 2.5D Effects}
DepthScape's 2.5D effects on static images can also be propagated to continuous image frames and create video 2.5D effects. Suggested by design experts, there can be two types of video creation. First, even in static images, the parameters of content placement can be bound to a time variable and rendered as animations. \textit{E.g.} a surrounding effect of an arrow ring, or a moving effect of a text plane. Second, continuous frames of videos can be parsed and edited with consistent parameters to enable object-centric video editing. Examples of both cases are shown in the video figure.

\subsection{Modify Real World Scenes}
\begin{figure}
    \centering
    \includegraphics[width=0.6\linewidth]{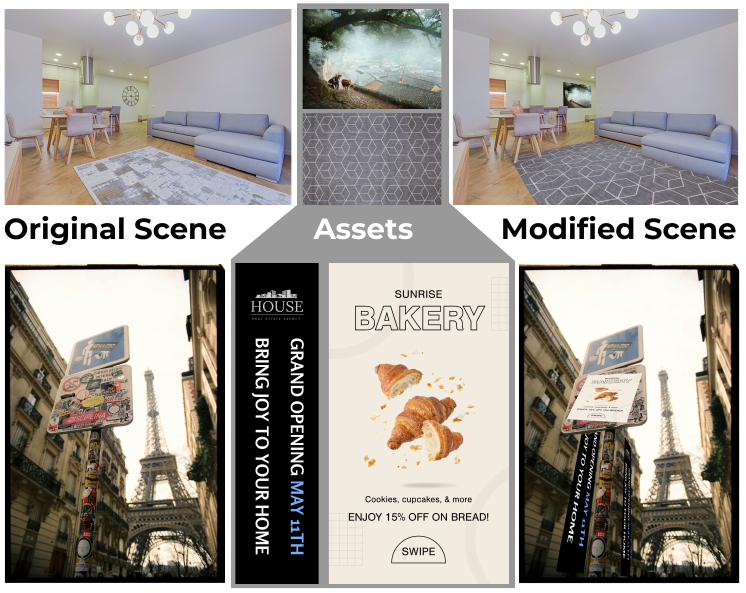}
    \caption{DepthScape can help adding contents into real world scenes. Left: original scenes; Middle: Design assets; Right: Modified scenes.}
    \label{fig:modify}
    \Description{A demonstration of DepthScape’s ability to integrate design assets into real-world scenes. The figure is divided into three columns: the left column shows two "Original Scene" images—an interior of a modern living room and a city street with the Eiffel Tower in the background. The middle column, labeled "Assets," displays four design elements: one art piece of a forest, another is a rug with a hexagonal pattern, and two vertical poster designs (a real estate advertisement and a bakery promotion). The right column, labeled "Modified Scene," shows the same original scenes with the assets inserted. In the living room, the hexagonal rug and forest wall art have been added. In the city street photo, the vertical signposts now include the real estate and bakery advertisements, aligned to match the perspective and depth of the scene.}
\end{figure}

Since DepthScape parses and approximates real-world depth cues and blends extra visual elements into the original scene, it can be useful to quickly modify real-world scenes in realistic ways, especially by adding new objects into the scenes. One intuitive example is simulating interior designs by adding new decorative elements, like image assets of rugs or paintings, into the scene; another example is testing branding visuals like posters by adding them to public spaces. See \autoref{fig:modify}. 

\subsection{Simulate AR Scenes}
\begin{figure}
    \centering
    \includegraphics[width=0.6\linewidth]{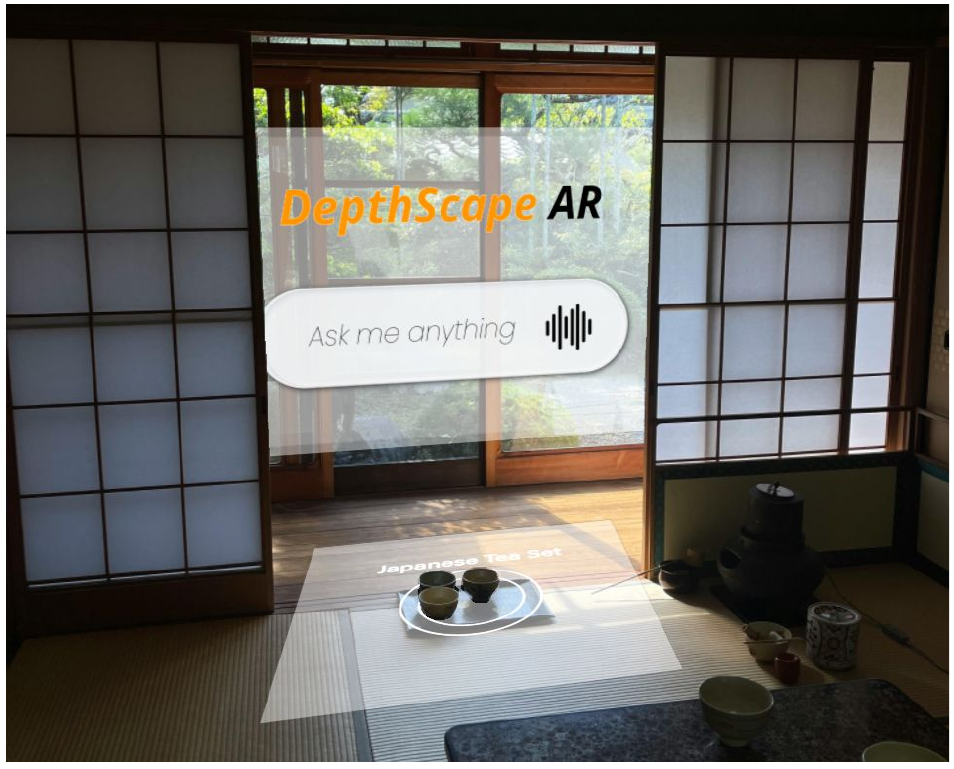}
    \caption{Simulated AR effect in a real-world scene.}
    \label{fig:AR}
    \Description{A simulated AR interface is overlaid on a photo of a traditional Japanese room with sliding paper doors and a view of a garden outside. The augmented reality overlay includes the label “DepthScape AR” at the top center, with a floating UI element below that says “Ask me anything” and a microphone icon, suggesting a voice interface. On the floor, near a low table with a tea set, there is a translucent panel with the label “Japanese Tea Set”, visually anchored to the real-world tea set, demonstrating how DepthScape can annotate physical objects in AR.}
\end{figure}

By anchoring contents to real-world objects and surfaces, DepthScape can also place and blend virtual assets, like UI and 3D models, into real-world images without actually tracking the environment. In this way, we are essentially simulating and prototyping AR effects. For example, by detecting the planar orientation of the floor and wall, we can blend AR contents into a real-world scene (\autoref{fig:AR}). We can also add animated 2D/3D contents, like a video or a 3D Pokémon, into real-world scenes.

\subsection{Storyboarding}
\begin{figure}
    \centering
    \includegraphics[width=0.5\linewidth]{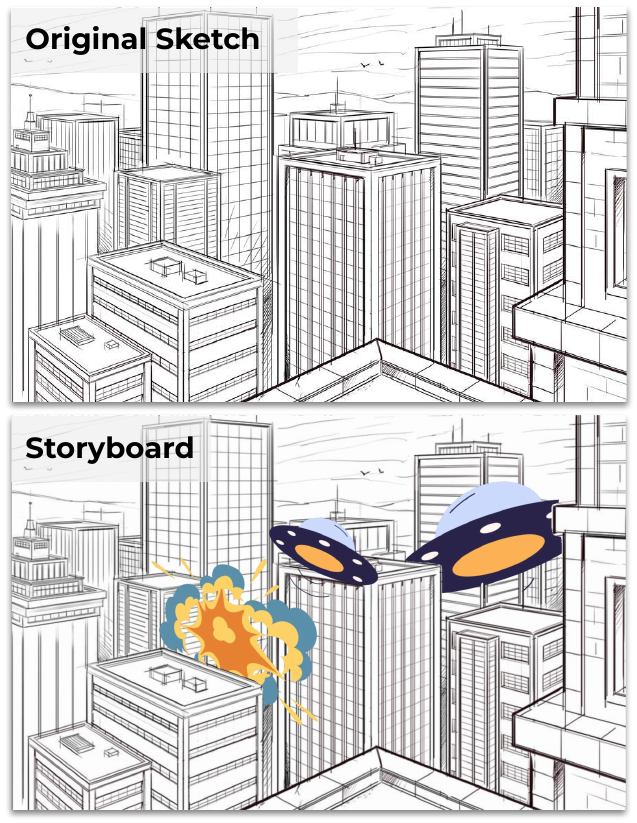}
    \caption{DepthScape can also apply to sketched scenes, helping users quickly story board their ideas.}
    \label{fig:storyboard}
    \Description{Two images vertically stackeds. The images show a cityscape sketch labeled "Original Sketch" and a modified version labeled "Storyboard." The "Original Sketch" is a black-and-white line drawing of tall buildings in an urban setting. The "Storyboard" version retains the background but includes colorful, comic-style elements: two flying UFOs in the sky and a large explosion among the buildings.}
\end{figure}

DepthScape can also be applied to hand-sketched scenes, enabling rapid blending of 2D contents for storyboarding. This capability supports content creators in visualizing and iterating on their designs more efficiently, allowing them to seamlessly integrate visual elements into sketched environments. In this case, DepthScape facilitates the blending of source materials with varying levels of fidelity, making it easier to unify assets ranging from rough sketches to polished graphics. See \autoref{fig:storyboard}.

\section{Discussion \& Future Work}
In this paper, we present \textit{DepthScape}, a human-AI collaborative authoring tool that enables 2D designers to create rich 2.5D visual effects by simply arranging assets within a \emph{monocular depth reconstruction} space derived from an input image.
A key strength of DepthScape is that it allows designers to work entirely on a 2D canvas to create diverse 2.5D effects with minimal 3D expertise.
DepthScape leverages an AI agent that orchestrates RGB–and–depth pipeline to analyze the input image and synthesize 3D geometry extraction code which, when compiled and executed, can generate content-aware parametric anchors that significantly simplify 2.5D editing and exploration.
We evaluated DepthScape through a multi-pronged study: a usability test with nine participants, technical performance on 100 stock images, expert validation with three professional designers, and exploratory demonstrations with five application scenarios.  

DepthScape occupies a distinctive middle ground in the design-tool landscape. Our evaluation revealed its intuitive usability and flexible editability. Designers can quickly manipulate 2D assets within a pseudo-3D space, while retaining the ability to edit with precision—bridging the gap between playful exploration and controlled refinement. On the contrary, when creating 2.5D visual designs, traditional 2D editors (\textit{e.g.,} Photoshop) demand tedious adjustments and strong spatial intuition. 3D software (\textit{e.g.,} Blender) automates perspective rendering but imposes steep learning curves and technical overhead. Generative editing tools (e.g., diffusion-based methods such as ``nano banana''\footnote{\href{https://aistudio.google.com/models/gemini-2-5-flash-image}{https://aistudio.google.com/models/gemini-2-5-flash-image}}) provide image editing capabilities but lack control and interactivity. Diffusion-based techniques also alter the entire original image, making them unsuitable when pixel-level fidelity must be preserved. DepthScape bridges these extremes by offering real-time, direct manipulation of 2D content in pseudo-3D space—retaining the quality of the original image while fostering playful exploration and serendipity. In this way, DepthScape complements both traditional editors and generative tools, providing an interactive, controllable, and exploratory pathway for depth-rich design.

\subsection{Semantics and Geometry Extraction}
Currently, DepthScape’s geometry extraction primarily handles single objects and simple geometric types. A promising direction for future work is to enable more complex forms of composition. For instance, combining the front-facing direction of a human figure with the normal of a ground plane could yield richer planar anchors. Our current visual program format supports such multi-element interactions, but realizing them will require additional examples data. Future deployments can help collect this data and expand the repertoire of supported geometries.

\subsection{Advanced Effects}
DepthScape enables realistic occlusion and perspective cues, but expert feedback points to opportunities for richer physical effects. Shadows, reflections on water or glass, color bleeding, fog, depth-of-field blur, and motion blur could make inserted content blend even more seamlessly into the scene. Incorporating these effects would require estimating additional environmental cues and upgrading the rendering pipeline, but such improvements could significantly extend DepthScape’s creative potential.


\subsection{Limitations}
Despite these contributions and future directions, several limitations remain in this paper.

\textbf{Evaluation design.} We did not conduct a direct comparison study against industry tools such as Photoshop or Blender. DepthScape is currently tailored to 2.5D effects rather than the broader image- and 3D-editing capabilities of these established platforms. At the same time, existing 2D, 3D, and generative approaches support different workflows and affordances, making it difficult to design fair and meaningful cross-platform tasks for comparative evaluation. Future work could develop task-specific benchmarks that highlight complementary strengths and situate DepthScape more clearly within the ecosystem of design tools.

\textbf{System performance.} Processing currently takes about 20 seconds per image, with the majority of time spent on waiting for VLM responses while other processing steps remain efficient. We expect processing times to decrease as newer and faster VLMs become available, and future work could also explore using smaller, task-specific VLMs and parallel execution of visual programs to further reduce latency.

\textbf{User interaction.} While our current design provides VLM-selected parametric anchors and intuitive direct manipulation editing, expert users desired finer controls—such as explicitly selecting target objects, specifying anchor types, or manipulating more sophisticated geometries (e.g., splines). We expect to extend the interaction options to incorporate such suggestions and better facilitate professional use.

\section{Conclusion}
Leveraging recent advancements in depth estimation and vision-language models, we present DepthScape, a novel graphic authoring tool for 2.5D design. DepthScape uses depth estimation to reconstruct a 3D scene from 2D input images, enabling the creation of realistic occlusion and perspective effects by placing and rendering visual elements within this 3D space. By incorporating vision-language models for semantic understanding of the scene, the system extracts parametric anchors from the reconstructed space, allowing users to place content through direct manipulation and parameter editing. We test the feasibility and usability of the system with a user study among both amateur and professional designers. The study results confirm the creativity benefits of DepthScape and guide the implementation of the final DepthScape interface. We further conduct a technical evaluation on 100 professional stock images, demonstrating efficiency, robustness, and versatility. Additionally, we validate the quality of DepthScape's outputs through an expert evaluation and five real-world application scenarios.

\bibliographystyle{ACM-Reference-Format}
\bibliography{ref}

\appendix

\appendix
\newpage
\section{Example Visual Programs}
\label{appendix:examples}
\textbf{Example1:}

The image depicts a draped figure covered entirely in a dark, flowing fabric, giving it an enigmatic and futuristic appearance. A circular digital text overlay with the words "4TH.CAPTERS" wraps around the figure, suggesting a technological or cyberpunk theme. 

Visual Program:

MASK\_0=Text2Mask(prompt = "the human figure")

POINTCLOUD\_0=Mask2Pointcloud(mask = MASK\_0)

CYLINDER\_0=Pointcloud2Cylinder(Pointcloud = POINTCLOUD\_0, direction = NULL)

CYLINDRICAL\_0=Cylindrical(cylinder = CYLINDER\_0)

\textbf{Example2:}

The image is a promotional poster for the "Kinosmena 2016" International Short Film Festival. It features a creative and modern design with a slightly side-facing woman's face framed by geometric panels that face the same direction, adorned with intricate blue foliage.

Visual Program:

MASK\_0=Text2Mask(prompt = "the human figure")

FACE\_0=FaceExtraction(mask = MASK\_0)

PLANAR=Planar(plane = FACE\_0.frontal)

\textbf{Example3:}

The image is a promotional poster for a "Senior Day" basketball event at Fifth Third Arena. It emphasizes the theme "Senior Day" with large white letters positioned prominently on the court, perpendicular to the ground and also in parallel to the square court's edge.

Visual Program:

MASK\_0=Text2Mask(prompt = "basketball playground")

POINTCLOUD\_0=Mask2Pointcloud(mask = MASK\_0) 

PLANE\_0=Pointcloud2PLANE(Pointcloud = POINTCLOUD\_0)

PLANAR=Planar(plane = PLANE\_0.extruded)

\textbf{Example4:}

The image is a poster for a "James Bond Symposium," prominently featuring a shiny gold bullet in a 3D perspective as the central visual element. Bold white text reading "JAMES BOND" wraps around the bullet, creating a dynamic and eye-catching effect.
Visual Program:

MASK\_0=Text2Mask(prompt="the bullet")

POINTCLOUD\_0=Mask2Pointcloud(mask=MASK\_0)

CYLINDER\_0=Pointcloud2Cylinder(Pointcloud=POINTCLOUD\_0, direction=NULL)

CYLINDRICAL\_0=Cylindrical(cylinder=CYLINDER\_0)

\textbf{Example5:}

The image depicts a draped figure covered entirely in a dark, flowing fabric, giving it an enigmatic and futuristic appearance. A spherical digital text overlay with the words "Darkness" wraps around the figure, suggesting a technological or cyberpunk theme. 

Visual Program:

MASK\_0=Text2Mask(prompt = "the human figure")

POINTCLOUD\_0=Mask2Pointcloud(mask = MASK\_0)

SPHERE\_0=Pointcloud2Sphere(Pointcloud = POINTCLOUD\_0)

SPHERICAL\_0=Spherical(sphere = SPHERE\_0)

\textbf{Example6:}

The image features a vintage-inspired design with a sepia-toned aerial view of a long, straight street cutting through an urban landscape, surrounded by shadowy buildings. The title "Invisible Streets" is prominently displayed along the street in bold, staggered white and yellow text, enhancing the sense of depth and direction.
Visual Program:

MASK\_0=Text2Mask(prompt = "highrise bridge in the input image")

POINTCLOUD\_0=Mask2Pointcloud(mask = MASK0)

PLANE\_0=Pointcloud2Plane(Pointcloud = Pointcloud0)

PLANAR=Planar(plane = PLANE\_0)

\textbf{Example7:}

The poster features a bold and energetic design, centered around a runner in motion. The background is white, with bright pink accents, including large, dynamic block text that forms an abstract, geometric pattern. This block text background is positioned to be in parallel to the running direction of the runner.

Visual Program:

MASK\_0=Text2Mask(prompt = "runner")

SKELETON\_0=SkeletonExtraction(mask=MASK\_0)

PLANAR\_0=Planar(plane = SKELETON\_0.median)

\textbf{Example8:}

The image showcases a vibrant aerial view of a city at night, filled with illuminated skyscrapers, streets, and bustling urban life. Overlaid on the cityscape, bold white typography spells out the phrase "NO OTHER GAME," cutting across the buildings with a dramatic, immersive perspective.

Visual Program:

MASK\_0=Text2Mask(prompt = "the front building in the input image")

POINTCLOUD\_0=Mask2Pointcloud(mask = MASK\_0)

PLANE\_0=Pointcloud2PLANE(Pointcloud = POINTCLOUD\_0)

PLANAR\_0=Planar(PLANE = PLANE\_0)

\textbf{Example9:}

The image features a relay race scene in a vibrant stadium setting, with one runner handing off the baton to another. The slogan "WE OFFER YOU THE BETTER, BETTER SOLUTION" is prominently displayed on the left side in bold, colorful text, placed in realistic perspective perpendicular to the ground but in parallel to the running direction of the runner.

Visual Program:

MASK0=Text2Mask(prompt = "ground")

Pointcloud0=Mask2Pointcloud(mask = MASK0)

MASK1=Text2Mask(prompt = "the runner in the middle")

SKELETON\_0=SkeletonExtraction(mask = MASK1)

PLANE\_0=Pointcloud2PLANE(Pointcloud = Pointcloud0)

PLANAR\_0=Planar(plane = SKELETON\_0.median)

\textbf{Example10:}

The image showcases two travelers walking towards an airport terminal. Both individuals are carrying luggage and backpacks, signaling they are either returning from or embarking on a journey. The text "BACK TO HOME" is prominently displayed in large white letters across the sky, with the perspective that looks like the text is parallel to the building facede.

Visual Program:

MASK\_0=Text2Mask(prompt = "building in the image")

POINTCLOUD\_0=Mask2Pointcloud(mask = MASK0)

PLANE\_0=Pointcloud2Plane(Pointcloud = Pointcloud0)

PLANAR\_0=Planar(plane = PLANE\_0)

\textbf{Example11:}

The image is a conceptual and artistic design featuring a person with their face pixelated and obscured by bright yellow text reading "\#31\#." The repeated hashtag emphasizes anonymity and digital communication themes. Surrounding the figure are additional yellow text and graphic elements, including "Mélancolique Anonyme" (Melancholic Anonymous), phone numbers, and technical details like "Appel non taxé" (Non-taxed call), evoking the idea of an anonymous support line or mental health service.

Visual Program:

MASK\_0=Text2Mask(prompt = "the human face")

FACE\_0=FaceExtraction(mask = MASK\_0)

POINTCLOUD\_0=Mask2Pointcloud(mask = MASK\_0)

SPHERE\_0=Pointcloud2Sphere(Pointcloud = POINTCLOUD\_0)

SPHERICAL\_0=Spherical(sphere = SPHERE\_0)

\textbf{Example12:}

The image is a minimalist and striking poster featuring a baseball player in mid-action, running against a solid bright blue background. Beneath the player, bold white text spells out "AUSTIN" in a stretched, perspective style, seemingly being placed on the ground and aligned with the facing direction of the player, adding depth and focus to the design.

Visual Program:

MASK0=Text2Mask(prompt ="floor")

Pointcloud0=Mask2Pointcloud(mask = MASK\_0)

MASK1=Text2Mask(prompt = "the player")

SKELETON\_0=SkeletonExtraction(mask = MASK\_1)

PLANE\_0=Pointcloud2PLANE(Pointcloud = POINTCLOUD\_0)

PLANAR\_0=Planar(plane = PLANE\_0)

\textbf{Example13:}

The poster promotes a 3D indoor biking experience with a Parisian theme. The Eiffel Tower dominates the background, and the tagline "VENHA PEDALAR PELAS RUAS DE PARIS" ("Come Pedal Through the Streets of Paris") is displayed in bold red and green text, emphasizing the immersive aspect of the activity. The text is placed perpendicular to the ground, while also in parallel to the edge of the road.

Visual Program:

MASK0=Text2Mask(prompt = "driveway")

Pointcloud0=Mask2Pointcloud(mask = MASK\_0) 

PLANE\_0=Pointcloud2PLANE(Pointcloud = POINTCLOUD\_0)

PLANAR\_0=Planar(plane = PLANE\_0.extruded)

\textbf{Example14:}

The poster is a bold, surreal design featuring a white classical bust with a futuristic twist. In the middle of the poster is a human head, with the top of the head appears cut open, revealing a black void, while the eyes glow red, giving a cyberpunk aesthetic. Large white typography dominates the composition, reading "WE ARE BACK IN BLACK", surrounding the head as a cylinder-like shape.

Visual Program:

MASK\_0=Text2Mask(prompt = "the human head")

FACE\_0=FaceExtraction(mask = MASK\_0)

POINTCLOUD\_0=Mask2Pointcloud(mask = MASK\_0)

CYLINDER\_0=Pointcloud2Cylinder(Pointcloud = POINTCLOUD\_0, direction = FACE\_0.cranial)

CYLINDRICAL\_0=Cylindrical(cylinder = CYLINDER\_0)

\textbf{Example15:}

The poster promotes a vocal training school for extreme vocal techniques, emphasizing styles like "Pig Voice," "Guttural," "Growling," and "Screaming," displayed in dynamic, radiating white text. The main visual element is a stylized blue-toned portrait of a person screaming, with lines and text emphasizing the sound's intensity and energy coming from her mouth.

Visual Program:

MASK\_0=Text2Mask(prompt = "the human figure")

FACE\_0=FaceExtraction(mask = MASK\_0)

PLANAR=Planar(plane = FACE\_0.median)

\textbf{Example16:}

The poster features a bold and energetic design, centered around a runner in motion. The background is white, with bright pink accents, including a large, dynamic block text that surrounds the runner. Since there isn't any visible ground surface in the poster, the cylinder of text is formed based on the runner figure's shape.

Visual Program:

MASK\_0=Text2Mask(prompt = "runner")

POINTCLOUD\_0=Mask2Pointcloud(mask = MASK\_0)

CYLINDER\_0=Pointcloud2Cylinder(Pointcloud = POINTCLOUD\_0, direction = NULL)

CYLINDRICAL\_0=Cylindrical(cylinder = CYLINDER\_0)

\end{document}